%% file: main.tex
\RequirePackage[l2tabu,orthodox]{nag}
\documentclass[submission,copyright]{eptcs}
\pdfoutput=1

\input{tikzpreamble.tex}

\usepackage{tikz}
\usepackage{placeins}
\usepackage{tabularx}
\usepackage{comment}

\input{preamble.tex}

\newcommand{\xx}{\ensuremath{\boldsymbol{x}}\xspace}
\newcommand{\yy}{\ensuremath{\boldsymbol{y}}\xspace}

\algdef{SE}[VARIABLES]{Variables}{EndVariables}
   {\algorithmicvariables}
   {\algorithmicend\ \algorithmicvariables}
\algnewcommand{\algorithmicvariables}{\textbf{global variables}}

\algblockdefx[Algorithm]{Algorithm}{EndAlgorithm}
    [1]{\textbf{algorithm} {\scshape #1}}
    {\textbf{end algorithm}}

\usetikzlibrary{matrix}

\usepackage{hyperref}
\hypersetup{
final=true,
draft=false
}

\title{Architecture-Aware Synthesis of Phase Polynomials for NISQ Devices}
\author{
\begin{tabular}{ccc}
Arianne Meijer-van de Griend$^{1,}$\thanks{ariannemeijer@gmail.com}
& & Ross Duncan$^{1,2,}$\thanks{ross.duncan@quantinuum.com} \\\\ 
\multicolumn{3}{c}{\footnotesize\makecell{$^1$ \textit{Quantinuum} \\ \textit{Terrington House, 13-15 Hills Road, Cambridge CB2 1NL, United Kingdom}}} \\
\multicolumn{3}{c}{\footnotesize\makecell{${}^2$ \textit{Department of Computer and Information Sciences}\\
\textit{University of Strathclyde}\\
\textit{26 Richmond Street, Glasgow, United Kingdom}}} \\
\end{tabular}
}
\def\titlerunning{Architecture-Aware Synthesis of Phase Polynomials for NISQ Devices}
\def\authorrunning{R.A. Meijer-van de Griend \& R. Duncan}

\begin{document}
\maketitle

\begin{abstract}
  We propose a new algorithm to synthesise quantum circuits for phase
  polynomials, which takes into account the qubit connectivity of the
  quantum computer.  We focus on the architectures of currently
  available NISQ devices.  Our algorithm generates circuits with a
  smaller \CNOT depth than the algorithms currently used in Staq and
  \tket, while improving the runtime with respect the former.
\end{abstract}

%
%
%


\section{Introduction}
\label{sec:intro}
Many current quantum computing architectures have restricted qubit
connectivity, meaning that interactions between qubits are only
possible when the physical qubits are adjacent in a certain graph,
henceforth called the \emph{architecture}, defined by the design of
the quantum hardware.  Traditional compiling techniques for quantum
circuits work around this limitation by inserting additional SWAP
gates into the circuit to move the logical qubits into a location
where the desired interaction is physically possible, a process called
\emph{routing} or \emph{mapping}
\cite{Alexander-Cowtan:2019aa,Soeken:2019aa,Zulehner:2017aa,Wille:2019aa}.
This typically increases the depth and gate count of the circuit by a
multiplicative factor between 1.5 and 3
\cite{Alexander-Cowtan:2019aa}.  However, recent work by Kissinger and
Meijer-van de Griend \cite{Kissinger:2019ac} has shown that for pure \CNOT
circuits it is possible to compile a circuit directly to an
architecture without dramatically increasing the number of \CNOT
gates. Their approach was to use a higher-level representation of the
desired unitary transform and (re)synthesise the corresponding circuit
in an architecture-aware manner.

In this paper, we consider another class of high-level constructs
called \emph{phase polynomials}, which give rise to circuits
containing only \CNOT and $R_Z(\theta)$ gates. 
The current state-of-the-art algorithm for phase polynomial synthesis is the
GraySynth algorithm \cite{Amy_2018}.  Unlike other algorithms for phase
polynomial synthesis \cite{Amy2014Polynomial-Time}, 
GraySynth attempts to minimise the number of 2-qubit gates.
 Unfortunately,
GraySynth assumes unrestricted qubit connectivity.  This
limitation was removed by Nash et al. \cite{Nash:2019aa}, by adding qubit
permutation subcircuits whenever a sequence of \CNOTs required by GraySynth is not
permitted by the architecture.  
Nevertheless, the algorithm still relies on the same recursive strategy as GraySynth, which might be suboptimal for sparse architectures.

In this paper we propose a new algorithm for the architecture-aware
synthesis of phase polynomial circuits.  The algorithm has been tuned
for the relatively sparse connectivity graphs of current quantum computers.  

We compare our algorithm against two compilers that are able to natively synthesise phase polynomials: \tket \cite{TKETPAPERHERE} and Staq \cite{staq}.
We compare the different methods based on the final \CNOT count, final
\CNOT depth, and their runtime.  These figures of merit are appropriate
for noisy-intermediate scale quantum (NISQ) devices
\cite{Preskill2018quantumcomputingin}, since the single-qubit gates of
such devices typically have error rates an order of magnitude less
than that of the two qubit gates.  By minimising the CNOT count we are
minimising the exposure of our computation to gate error, including crosstalk; by
minimising depth we reduce its exposure to decoherence.


We show that for sufficiently sparse quantum computer architectures
and sufficiently large phase polynomials, our algorithm outperforms
the algorithm from Nash et al.~\cite{Nash:2019aa} that is used in Staq
\cite{staq} as well as the decomposition and routing strategies from
\tket \cite{Alexander-Cowtan:2019aa}.  Our algorithm relies on finding
non-cutting vertices in the connectivity graph, and does not require
computing any Steiner trees; we find that in most cases our algorithm
has reduced runtime compared to that of Nash et al.

In section \ref{sec:related}, we introduce phase polynomials and
existing methods for their synthesis, both with and without
architecture-awareness.  Our new algorithm is described in section
\ref{sec:methods} and our experimental results can be found in section
\ref{sec:results}.   Throughout the paper we will assume some
familiarity with the \zxcalculus \cite{Coecke:2009aa}, which we use as
notation.   For the uninitiated, Cowtan et al.~\cite{Cowtan:2019aa} give a short
introduction to the calculus, including the phase gadget notation;
Coecke and Kissinger provide a complete treatment
\cite{Coecke2017Picturing-Quant}.  

\paragraph*{Notation}
\label{sec:notation}
We use bold face letters $\xx$, $\yy$, to
denote vectors, and the corresponding regular weight letters $x_i$,
$y_j$ to denote their components.

\section{Phase polynomial synthesis}
\label{sec:related}

Following Amy et al.~\cite{Amy_2018}, we define the phase polynomial
via the sum-over-paths formalism \cite{Dawson:2005aa}.
\begin{definition}
  Let $C$ be a circuit consisting of only \CNOT and $R_{Z}(\theta)$ gates; then
  its corresponding unitary matrix $U_C$ has a \emph{sum-over-paths} form,
\begin{equation}\label{eq:sum-over-paths}
U_C = \sum_{\xx \in \mathbb{F}_2^n} e^{2 \pi i f(\xx)}\ket{A\xx}\bra{\xx}
\end{equation}
consisting of a \emph{phase polynomial} 
\begin{equation}\label{eq:phasepoly}
  f(\xx) = \sum_{\yy \in \mathbb{F}_2^n} \hat{f}(\yy) \cdot (x_1y_1 \oplus x_2y_2 \oplus \cdots \oplus x_ny_n)
\end{equation}
with \emph{Fourier coefficients} $\hat{f}(\yy) \in \mathbb{R}$, and a
\emph{basis transform} $A \in GL(n, \mathbb{F}_2)$.  When no confusion
will arise we refer to the pair $(f, A)$ as the phase polynomial of
$C$.
\end{definition}

\noindent
Note that parity functions -- henceforth just called \emph{parities}
-- of the form $\xx \mapsto (x_1y_1 \oplus \cdots \oplus x_ny_n)$ as
in Equation \ref{eq:phasepoly}, can be identified with the bit string
\yy; these are the basis of the space of phase polynomials.  Those
parities for which $\hat f(\yy) \neq 0 $ are called the \emph{support}
of $f$.

Every circuit over $\{\CNOT, R_Z(\theta)\}$ has a canonical sum-over-paths
form, which we now sketch.  First, we associate a parity to each ``wire
segment'' of the circuit as follows: the inputs of the
circuit are labelled $x_1,\ldots,x_n$ respectively; the output of an
$R_Z$ gate has the same parity as its input; and a \CNOT gate with
parities $p_1$ and $p_2$ on its control and target inputs has output
parities $p_1$ and $p_1 \oplus p_2$, respectively.  Second, the
coefficients $\hat f(\yy)$ are computed by summing all the angles $\theta$
occurring in $R_Z$ gates labelled by the parity \yy.  Finally, the
linear transform $A$ is defined by the mapping $\xx \mapsto \xx'$
where $\xx'$ are the final labels of circuit outputs.  We refer the
reader to Amy et al.~\cite{Amy_2018} for more details. 

The task of \emph{phase polynomial synthesis} is the reverse: given
$(f,A)$ we must find the circuit $C$.  This amounts to constructing a parity
labelled \CNOT circuit such that every \yy in the support of $f$
occurs as a label on some wire, adding an $R_Z(\hat f (\yy))$ gate on
that wire, and extending that circuit so that the desired output
parities for $A$ are achieved.  Since $f(\xx)$ is a sum, and
addition is commutative, the order in
which the parities are achieved is irrelevant; neither does it matter on
which qubits these parities occur.
To obtain the required final parities, additional \CNOTs are added to
the circuit.  Since the new parity is the sum of the parities of
both the control and the target qubit, applying a \CNOT gate can
therefore be seen as an elementary row operation on the matrix $\xx
\mapsto \xx'$.  If the desired
parities for each qubit are known, Gaussian elimination can produce a
\CNOT sequence to achieve those parities \cite{patel, Kissinger:2019ac,
Nash:2019aa}. This method suffices to synthesise the matrix $A$ of the
phase polynomial \cite{Amy2014Polynomial-Time,Amy_2018, Nash:2019aa};
note, however, that this second phase is totally independent of the 
earlier synthesis of the parities required for $f(\xx)$.

\paragraph{Architecture agnostic synthesis.}

Phase polynomials may be synthesised via the \emph{phase gadget}
construct of the \zxcalculus  \cite{Cowtan:2019aa}.  
Since our algorithm can be intuitively described using phase gadgets,
we will briefly explain this method.

\begin{definition}\label{def:phasegadget}
  In \zxcalculus notation we denote the $R_Z$ gate with phase $\alpha$ and \CNOT gate as:
  \[
    {\scalebox{0.65}{\inltf{GatesRZ}}} \simeq {\scalebox{0.65}{\inltf{ZXCalcRZ}}} \quad \quad \quad
    {\scalebox{0.65}{\inltf{GatesCX}}} \simeq {\scalebox{0.65}{\inltf{ZXCalcCX}}}
    \]
  In a phase polynomial $(f, A)$, each term in $f(\xx)$  defines an
  operator $e^{-i\frac{\alpha}{2}Z^{\otimes n}}$, which we represent
  by the \emph{phase gadget} $\Phi_n(\alpha)$ : 
    \[
  \Phi_n(\alpha) := {\scalebox{0.65}{\inltf{PhaseGadgetDef}}}
  \]
  where $\alpha = \hat{f}(\yy)$ and the gadget is connected to qubit
  $i$ iff $x_iy_i = 1$.
\end{definition}


\begin{lemma}\label{lem:basic-phsgad-eqns}
  We have the following law for decomposition of phase gadgets \cite{Cowtan:2019aa}.\\
  \begin{tabularx}{\textwidth}{Xp{1cm}X}
    \vspace{-20pt} 
\begin{equation}\label{eq:place-cnot}
  {\scalebox{0.65}{\inltf{PhaseGadgetCNOT-red1}}}
  = {\scalebox{0.65}{\inltf{PhaseGadgetCNOT-red3}}}
\end{equation} & & 
\begin{equation}\label{eq:place-phase}
  {\scalebox{0.65}{\inltf{PhaseGadgetPhase-lhs}}}
  = {\scalebox{0.65}{\inltf{PhaseGadgetPhase}}}
\end{equation}
  \end{tabularx}
\end{lemma}

Lemma \ref{lem:basic-phsgad-eqns} serves as a recursive definition of
the phase gadget, and demonstrates how the gadget may be realised as
two ladders of \CNOTs and an $R_Z$ gate.  Cowtan et
al.~\cite{Cowtan:2019aa} showed how to synthesise phase gadgets in
reduced depth using a balanced tree of \CNOTs, however if the gadgets
are synthesised singly, and their ordering is not taken into account,
the circuit may still be suboptimal even after local optimisation.

A consequence of Lemma \ref{lem:basic-phsgad-eqns} is that phase
gadgets stabilise \CNOT circuits in the following sense.  Let $C_{ij}$
be a \CNOT gate with control qubit $i$ and target qubit $j$; then for
all phase gadgets $\Phi(\alpha)$ there exists $\Phi'(\alpha)$ such that
$C_{ij} \Phi(\alpha) = \Phi'(\alpha) C_{ij}$.  $\Phi'$ is identical to
$\Phi$ except that $\Phi'$ is connected qubit $i$ iff $\Phi$ is
connected to exactly one of $i$ and $j$.

This observation leads to an improvement in the algorithm.  If we view
the sequence of phase gadgets as a binary matrix whose rows are the
qubits and whose columns are the
corresponding parities \yy in the support of $f$, then commuting
$C_{ij}$ through the entire circuit is an elementary row operation,
namely adding row $j$ to row $i$.  Therefore, by conjugating the
circuit with \CNOTs, we may obtain a column containing a single 1.  At
that point, the desired parity (corresponding to the column in the
matrix) is achieved on the qubit corresponding to the row with the
1. The $R_Z$ gate can then be placed, and the column can be removed
from the matrix.

For example, the 3 qubit phase polynomial, ($f(\xx)$, $I$), specified by 
$
f(\xx) = \alpha_1 (x_2 \oplus x_3) + \alpha_2 (x_1 \oplus x_2) + \alpha_3 (x_1 \oplus x_3) + \alpha_4 x_3,
$
can be represented in a \zxdiagram and corresponding binary matrix as:
\[
{\scalebox{0.65}{\inltf{3qExample}}} \quad \sim \quad  
\begin{pmatrix}
  0 & 1 & 1 & 0 \\
  1 & 1 & 0 & 0 \\
  1 & 0 & 1 & 1
\end{pmatrix}
\]
Conjugating the first and second qubits with two \CNOTs, and applying
Eq.~\ref{eq:place-cnot} we obtain the following rewrite sequence and
final matrix:
\[
  {\scalebox{0.65}{\inltf{3qExample}}}  = 
  {\scalebox{0.65}{\inltf{3qExample1}}} =
  {\scalebox{0.65}{\inltf{3qExample2}}} =
  {\scalebox{0.65}{\inltf{3qExample3}}} =
\]
\[
  {\scalebox{0.65}{\inltf{3qExample4}}} =
  {\scalebox{0.65}{\inltf{3qExample5}}} \sim
\begin{pmatrix}
  1 & 0 & 1 & 0 \\
  1 & 1 & 0 & 0 \\
  1 & 0 & 1 & 1
\end{pmatrix}
\]
The second and last columns of the matrix contain only a single $1$,
so we can use Equation \ref{eq:place-phase} to place a $R_Z$ gate:
\[
  {\scalebox{0.65}{\inltf{3qExample5}}} \quad = \quad 
  {\scalebox{0.65}{\inltf{3qExample6}}} \quad = \quad 
{\scalebox{0.65}{\inltf{3qExample7}}}   \quad \sim \quad  
\begin{pmatrix}
  1 & 1 \\
  1 & 0 \\
  1 & 1
\end{pmatrix}
\]
Note the equation relies on the fact that $R_Z$ gates commute with
phase gadgets.


The matrix representation reduces the task of phase polynomial synthesis to
finding the order in which to reduce the columns,  and which qubit
should remain a 1 in the matrix for each column. Amy et
al.~\cite{Amy_2018} proposed a heuristic algorithm called
\emph{GraySynth} based on Gray codes. 
The main idea is to pick the qubit $q$ participating in most parities and 
then achieving all parities containing $q$ in order of 
Gray codes \cite{gray} on qubit $q$.
As a result, many \CNOTs will have the same target qubit.
This algorithm has been implemented as part of Staq \cite{staq} in combination with SWAP-based routing.


Unfortunately, GraySynth does not accommodate qubit connectivity
restrictions,  making it less useful for NISQ devices.
A naive solution is to apply a generic qubit routing routine to the
synthesised circuit, 
however this will almost always increase the size of the circuit.
Luckily, there is no need to be so naive.

\paragraph{Architecture-aware synthesis.}
It is possible to define synthesis algorithms which produce
circuits that immediately satisfy the constraints imposed by the quantum computer. 
Several algorithms such \emph{architecture-aware synthesis} algorithms
 for \CNOT circuits and phase polynomials have 
recently been proposed \cite{Kissinger:2019ac, Nash:2019aa}. 
While SWAP-based methods respect the original structure of the circuit at
the level of individual gates, architecture-aware synthesis preserves
only the overall unitary, and this additional freedom allows the
architectural constraints to inform the choice of which gates to generate.
This concept has also been used in the Staq compiler \cite{staq}, 
which uses the algorithms described in this section.

Kissinger et al.~\cite{Kissinger:2019ac} and Nash et
al.~\cite{Nash:2019aa} independently modified the Gaussian
elimination algorithm sketched above to synthesise routed \CNOT circuits.
They used Steiner trees to determine paths on the connectivity graph across 
which to simulate one or more \CNOT gates. 
Nash et al.~\cite{Nash:2019aa} showed that their method  
scales well with respect to the size and the density of the connectivity graph of the quantum computer. 
Kissinger et al.~\cite{Kissinger:2019ac} showed that for circuits consisting only of \CNOT gates 
their method outperformed current state-of-the-art SWAP-based methods. 
Wu et al. \cite{wu2019optimization} have recently improved these
methods with an adaptation relying on Steiner trees and non-cutting vertices.

This constrained version of Gaussian elimination, called \emph{Steiner-Gauss}, 
can be used in any synthesis algorithm by replacing the original Gaussian 
elimination such that it routes (part of) the synthesised circuit. 
In particular, this can be used in the T-par algorithm \cite{Amy2014Polynomial-Time} and 
in GraySynth it can be used to synthesise the matrix $A$.

Nash et al. \cite{Nash:2019aa} also proposed an adaptation of the GraySynth algorithm 
we called \emph{Steiner-GraySynth}. 
They replaced the step in the original GraySynth algorithm 
that generates a small sequence of \CNOTs with a step 
that emulates this sequence with routed \CNOTs. 
This emulation is created using a Steiner tree over the connectivity graph 
with the phase qubit as root and 
the other qubits participating in the sequence of \CNOTs as nodes. 
Then, a \CNOT is placed for every Steiner-node in the tree 
and one for every edge in the Steiner tree.

For phase polynomial synthesis, this algorithm performs better than
naive routing \cite{Nash:2019aa}.  However, following GraySynth, it
will place many \CNOT gates with the same target qubit.  If this qubit
is poorly connected in the architecture, a large \CNOT overhead will
result.  Furthermore, it requires the construction of a Steiner tree
in order to route the \CNOT gates.  The minimal Steiner tree problem
is NP-hard\cite{steinercompl}, so finding the true optimum is not
feasible, but it can be approximated in polynomial time using the
all-pairs shortest paths and building a spanning tree between them.  

\section{New natively routed heuristic algorithm}
\label{sec:methods}

\begin{figure}[t!]
  \begin{algorithmic}
\Variables
  \State $G$, the architecture graph
  \State $Circuit$, An initially empty circuit with $|G.\mathrm{vertices}|$ qubits
  \State $A$, The basis transform of the phase polynomial
  \State $P$, The matrix describing the support of $f$
  \State $ZPhases$, The list of Z phases $\hat f(\yy)$ belonging to each parity \yy in $f$
\EndVariables
\State
\Function{BaseRecursionStep}{$Cols$, $Qubits$}
\If {$Qubits$ non-empty and $Cols$ non-empty}
  \State $H \gets $ InducedSubgraph($G$,$Qubits$) 
  \State $Rows \gets $ NonCuttingVertices($H$) 
  \State $ChosenRow \gets \argmax_{r \in Rows} \,
                          \max_{x\in \mathbb{F}_2} \, 
                          |\{c\in Cols \text{ where } P_{r,c} = x\}| $
  \State $Cols0, Cols1 \gets $ SplitColsOnRow($Cols$, $ChosenRow$)
  \State BaseRecursionStep($Cols0$, $Qubits \setminus \{ChosenRow\}$)
  \State OnesRecursionStep($Cols1$, $Qubits$, $ChosenRow$)
\EndIf
\EndFunction
\State
\Function{OnesRecursionStep}{$Cols$, $Qubits$, $ChosenRow$}
  \If {$Cols$ non-empty}
    \State $Neighbours \gets \{q \in Qubits \text{ where } q \sim
    ChosenRow \text{ in } G \}$
    \State $n \gets \argmax_{q \in Neighbours} 
                    |\{c\in Cols \text{ where } P_{q,c} = 1\}| $
    \If{$|\{c\in Cols \text{ where } P_{n,c} = 1\}| > 0$}
      \State Place\CNOT($ChosenRow$, $n$)
      \State $Cols \gets $ ReduceColumns($Cols$)
    \Else
      \State Place\CNOT($n$, $ChosenRow$)
      \State Place\CNOT($ChosenRow$, $n$)
    \EndIf
    \State $Cols0, Cols1 \gets $ SplitColsOnRow($Cols$, $ChosenRow$)
    \State BaseRecursionStep($Cols0$, $Qubits \setminus \{ChosenRow\}$)
    \State OnesRecursionStep($Cols1$, $Qubits$, $ChosenRow$)
  \EndIf
\EndFunction
\State
\Algorithm{RoutedPhasePolySynth}
    \State $Columns \gets $ ReduceColumns($\{0,\ldots, \,|P.\mathrm{columns}|\}$)  
    \State BaseRecursionStep($Columns$, $G.\mathrm{vertices}$)
    \State $Circuit$.AddGates(SteinerGauss($A*P'^{-1}$))
    \EndAlgorithm
  \end{algorithmic}
  \caption{Algorithm for synthesising phase polynomials in an
    architecture aware manner.  The subroutines not defined here are
    in listed in Appendix \ref{apx:sub_alg}.}
  \label{fig:main_alg}
  \figureline
\end{figure}

In this section, we describe a natively routed algorithm that attempts
to take the architecture into account. It uses a novel heuristic
which works well for sparse architecture graphs.

Pseudo-code for the algorithm is shown in Figure
\ref{fig:main_alg} and its sub-procedures are listed in Appendix
\ref{apx:sub_alg}.  
%
A full worked example is presented in
Appendix \ref{apx:example}; for ease of comparison this example is the
same one treated by Amy et al. \cite{Amy_2018} using the GraySynth algorithm.

In the following, the architecture graph -- that is, the connectivity
map of the physical qubits -- is denoted $G$.  The phase polynomial to
be synthesised, $(f,A)$, is represented as two binary matrices, $P$
and $A$, where the columns of $P$ are the corresponding parities \yy
in the support of $f$, as explained in Section \ref{sec:related}.  By
construction, the columns in $P$ are unique and no column $\yy$ has
all values set to $0$.

\paragraph{Preprocessing.}
The algorithm starts by synthesising phase gadgets of the form specified by Equation \ref{eq:place-phase}. 
This will remove trivial columns in $P$ and 
placing their corresponding $R_Z$ phase gates. 
A column \yy is trivial if it has exactly one index $j$ such that $y_j = 1$. 
The phase gate $R_Z$ is placed on the qubit corresponding to $j$ and 
its phase $\alpha$ is equal to $\hat{f}(\yy)$.
This makes sure that every column $\yy$ in $P$ contains at least two elements with value $1$. 
Hence, each column requires at least one \CNOT in order to be synthesised by Lemma \ref{lem:basic-phsgad-eqns}.

For example, consider the phase polynomial from Section \ref{sec:related}, 
we can use Equation \ref{eq:place-phase} to remove the fourth column 
(corresponding to $\alpha_4$) and synthesise the phase gate $R_Z(\alpha_4)$ on qubit $3$:
\[
  \begin{pmatrix}
    0 & 1 & 1 & 0 \\
    1 & 1 & 0 & 0 \\
    1 & 0 & 1 & 1
  \end{pmatrix} \sim   
  {\scalebox{0.65}{\inltf{3qExample}}} =
{\scalebox{0.65}{\inltf{3qExample8}}} \sim
\begin{pmatrix}
  0 & 1 & 1 \\
  1 & 1 & 0 \\
  1 & 0 & 1 
\end{pmatrix}
\]

\paragraph{Base recursion step.}
Similar to GraySynth, we want to synthesise the phase gadgets in the phase polynomial in an order that requires the least amount of \CNOT gates.
However, we do not want to synthesise the phase gadgets such that many phase gates are placed on the same qubit.
Instead, we pick one qubit and attempt to remove its row from $P$.
However, we cannot pick just any row to remove from $P$ because it might still be needed to synthesise other phase gadgets due to the connectivity constraints.
Thus, we pick a non-cutting vertex $i \in G$ such that row $P_i$ has either the most ones or the most zeroes. 
A \emph{non-cutting vertex} is a vertex in $G$ that can be removed
from $G$ without disconnecting the remaining graph.
Like GraySynth, we split $P$ into two matrices, $P^0$ and $P^1$, 
such that column $P_j$ is a column in $P^0$ iff $P_{i,j}=0$ and $P_j$ is a column in $P^1$ otherwise. 
Since all entries in row $P^0_i$ are equal to $0$, we do not need this
row any more and we can remove it from $P^0$, and because $i$ is
non-cutting, its removal leaves the graph connected.  Then, we use the
base recursion step on the sub-matrix $P^0$ (excluding row $i$) with
the sub-graph of $G$ where vertex $i$ has been removed. 
The matrix $P^1$ is treated by a different recursive procedure using
the full graph $G$, described below.

Continuing the example above, suppose we are targeting the
architecture $G : x_1 \Leftrightarrow x_2 \Leftrightarrow x_3$.  We
can pick either $x_1$ or $x_3$ as they are both non-cutting and have the same number of ones
and zeroes; we will make the arbitrary choice of $x_1$. This choice
yields our new $P^0$ and $P^1$:
\[
P = 
  \begin{pmatrix}
    0 & 1 & 1 \\
    1 & 1 & 0 \\
    1 & 0 & 1 
  \end{pmatrix}
\qquad 
P^0 = 
  \begin{pmatrix}
    1 \\
    1 
  \end{pmatrix}
\qquad
P^1 = 
  \begin{pmatrix}
    1 & 1 \\
    1 & 0 \\
    0 & 1 
  \end{pmatrix}
\]
Note that $P^0$ corresponds to the phase gadget $\alpha_1$, and $P^1$
corresponds to the phase gadgets $\alpha_2$ and $\alpha_3$.  Recursing
on $P^0$ will eventually place the \CNOT $C_{3,2}$ and $R_Z(\alpha_1)$
gate on qubit $x_2$, as shown below.
\[
  \begin{pmatrix}
    0 & 1 & 1 \\
    1 & 1 & 0 \\
    1 & 0 & 1 
  \end{pmatrix} \sim   
  {\scalebox{0.65}{\inltf{3qExample8}}} =
  {\scalebox{0.65}{\inltf{3qExample9}}} =
  {\scalebox{0.65}{\inltf{3qExample10}}} \sim
\begin{pmatrix}
  1 & 1 \\
  1 & 0 \\
  1 & 1 
\end{pmatrix}
\]

Bear in mind that the recursion on $P^0$ may add \CNOTs to the circuit,
performing a row operation on the global $P$ matrix.  For our
recursion scheme to be valid we require that the row $P^1_i$ remains
equal to $1$.  Initially, this holds by the construction of $P^1$.
Since row $i$ has been removed from $P^0$, no gate involving qubit $i$
will be added by recursion on $P^0$, and hence the $i$th row of $P^1$
will be unchanged.
%
%
Moreover, $P^1$ does not contain any trivial columns
because for every column $j$ in $P^1$ there will always be another row $k \neq i$ such that $P^1_{k,j} = 1$. 

\paragraph{Ones recursion step.}

The recursion step for $P^1$ attempts to remove as many ones 
from row $i$ as possible such that it can be removed.
This can be achieved by placing \CNOTs in the circuit, however we are 
restricted by the connectivity graph.
Therefore, we pick a neighbour vertex $n \in G$ such that row $P^1_n$ has most ones.
Picking the row $P^1_n$ with most ones will ensure that most ones are removed.
Then, we can conjugate with \CNOT $C_{i,n}$, and update $P$ by adding
row $n$ to row $i$, as explained in Section \ref{sec:related}.
This might introduce trivial columns in $P^1$ 
(note that $P^0 = \emptyset$ by the recursion), 
which are removed like in the preprocessing step.
Thus, in the example circuit:
\[
\begin{pmatrix}
  1 & 1 \\
  1 & 0 \\
  1 & 1 
\end{pmatrix}
\sim
  {\scalebox{0.65}{\inltf{3qExample10}}} =
  {\scalebox{0.65}{\inltf{3qExample11}}}  
\sim
\begin{pmatrix}
  0 & 1 \\
  1 & 0 \\
  1 & 1 
\end{pmatrix}
\]

However, if every entry of row $P^1_n$ is $0$, conjugation with
$C_{i,n}$ will have no effect.  In this situation, we first apply the
opposite \CNOT,  $C_{n,i}$ and then $C_{i,n}$ as before.  This
effectively swaps the rows $i$ and $n$, so there is no need to reduce
the circuit. Nevertheless, this ensures that every entry in row $P^1_i$ is $0$.

After placing the \CNOT gate(s), we have modified row $P^1_i$ and 
we can split $P^1$ into two matrices, $P^{1,0}$ and
$P^{1,1}$, and recurse upon these two as in the base recursion step.

In our example $P^{1,0}\sim \{\alpha_2\}$ and $P^{1,1}\sim \{\alpha_3\}$. 
We use the base recursion on $P^{1,0}$ and pick $x_3$ arbitrarily. 
Note that we only consider the sub-matrix $P^{1,0}$ to count the number of ones and zeroes. 
Then, we split $P^{1,0}$ into $\emptyset$ and $\{\alpha_2\}$, respectively. 
In the ones recursion step, we pick neighbour $x_2$ and 
place \CNOT $C_{3,2}$ and $R_Z(\alpha_2)$ on qubit $x_2$.
\[  
  {\scalebox{0.65}{\inltf{3qExample11}}} =
  {\scalebox{0.65}{\inltf{3qExample12}}} =
  {\scalebox{0.65}{\inltf{3qExample13}}} \sim
\begin{pmatrix}
  1 \\
  1 \\
  1 
\end{pmatrix}
\]
Afterwards, we use the ones recursion step twice on the remaining row, 
placing two \CNOTs, $C_{2,1}$ and $C_{3,2}$, and placing the final phase gate $R_Z(\alpha_3)$ on qubit $x_3$ 
\[  
  {\scalebox{0.65}{\inltf{3qExample14}}} =
  {\scalebox{0.65}{\inltf{3qExample15}}} =
  {\scalebox{0.65}{\inltf{3qExample16}}} 
%
\]

\paragraph{Post-processing.}

Lastly, we need to synthesise the basis transform $A$.
Because the \CNOT gates in the circuit, obtained by synthesising the phase gadgets, change the parities on each qubit, we need to undo these changes.
Let $P'$ be the basis transform corresponding to the final parities of the synthesised circuit, 
then we can undo this transform and apply the desired transform $A$ 
by synthesising $A\cdot P'^{-1}$ using Steiner-Gauss as explained in Section \ref{sec:related}. 

In our synthesis example, $P'^{-1}$ is equivalent to the \CNOTs that were commuted 
to the end of the \zxdiagram. 
Incidentally, $P'^{-1} = A\cdot P'^{-1}$ because of our choice $A=I$, 
thus the desired linear transformation is already achieved. 
Moreover, these trailing \CNOTs are 
already routed, however resynthesising them might remove a few 
redundant \CNOTs for the final circuit. 

\paragraph{Termination and correctness.}

Our algorithm terminates and is correct if the recursion converges and 
it synthesises the desired phase polynomial $(f, A)$ while satisfying 
the connectivity constraints imposed by the architecture.

At each recursion step, the matrix $P$ is split into $P^0$ and $P^1$.
In the case of $P^0$, the base recursion step will effectively remove a row from $P^0$.
In the case of $P^1$, the ones recursion step will place one or two \CNOT gates. 
This will either remove a column from $P^1$ or, when splitting $P^1$ into $P^{1,0}$ and $P^{1,1}$, make sure that $P^{1,0} \neq \emptyset$.
The recursion finishes when $P$ is empty.
Hence, the recursion converges and the algorithm terminates.

By construction, the matrix $P$ describes the remaining phase gadgets to be synthesised 
(initially the parities \yy in the support of $f$).
This remains the case while synthesising because placing a \CNOT updates $P$ 
with an elementary row addition as explained in Section \ref{sec:related}.
Moreover, a column is only removed from $P$ iff the phase gadget is trivial, 
i.e. it is of the form described by Equation \ref{eq:place-phase}.
Consequently, the phase gates are placed at the right parity by Lemma \ref{lem:basic-phsgad-eqns}.
Lastly, the basis transform $A$ is obtained as described in the previous paragraph.
Thus, the algorithm has synthesised the desired phase polynomial once it has terminated.


Additionally, all \CNOT gates that are added have the property that 
the control and target qubits are neighbours in the connectivity graph $G$, thus satisfying the connectivity constraints imposed by the architecture.

Hence, our algorithm terminates and when it does the desired phase polynomial has been synthesised in an architecture-aware manner.


\FloatBarrier
\section{Results and discussion}
\label{sec:results}
\begin{figure}[t]
  \centering
  \begin{subfigure}[b]{\textwidth}
  \includegraphics[width=\textwidth]{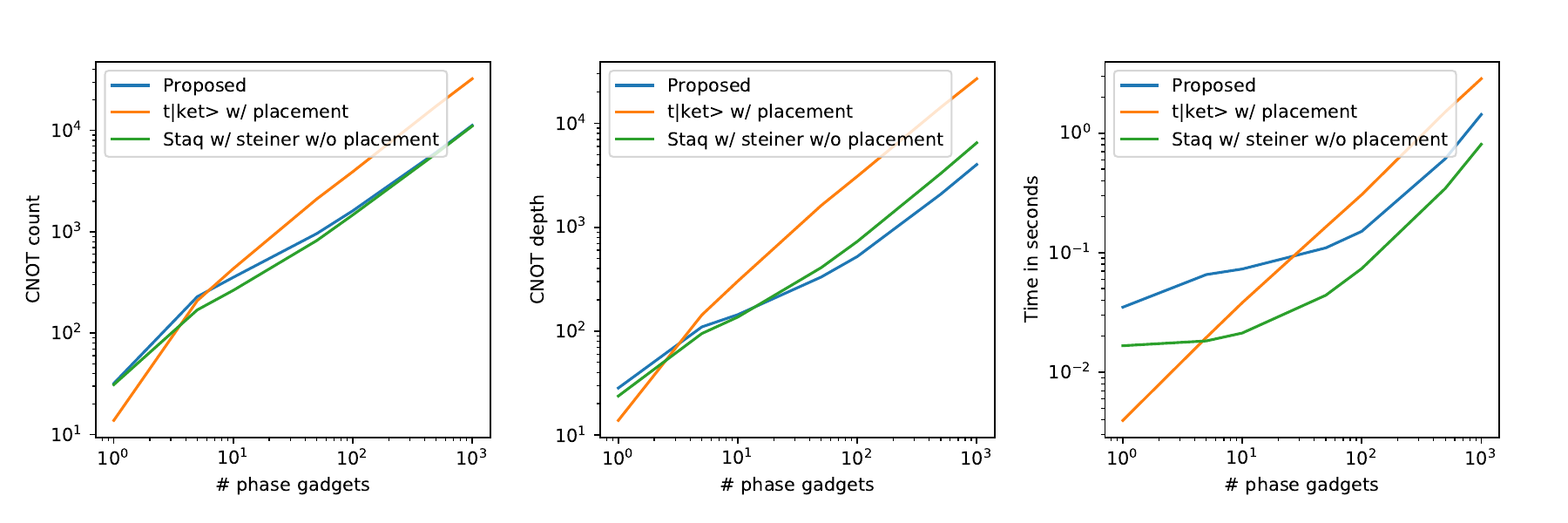}
  \subcaption{16 qubit Rigetti Aspen}\label{fig:gadgetscale_aspen}
  \end{subfigure}
  \begin{subfigure}[b]{\textwidth}
  \includegraphics[width=\textwidth]{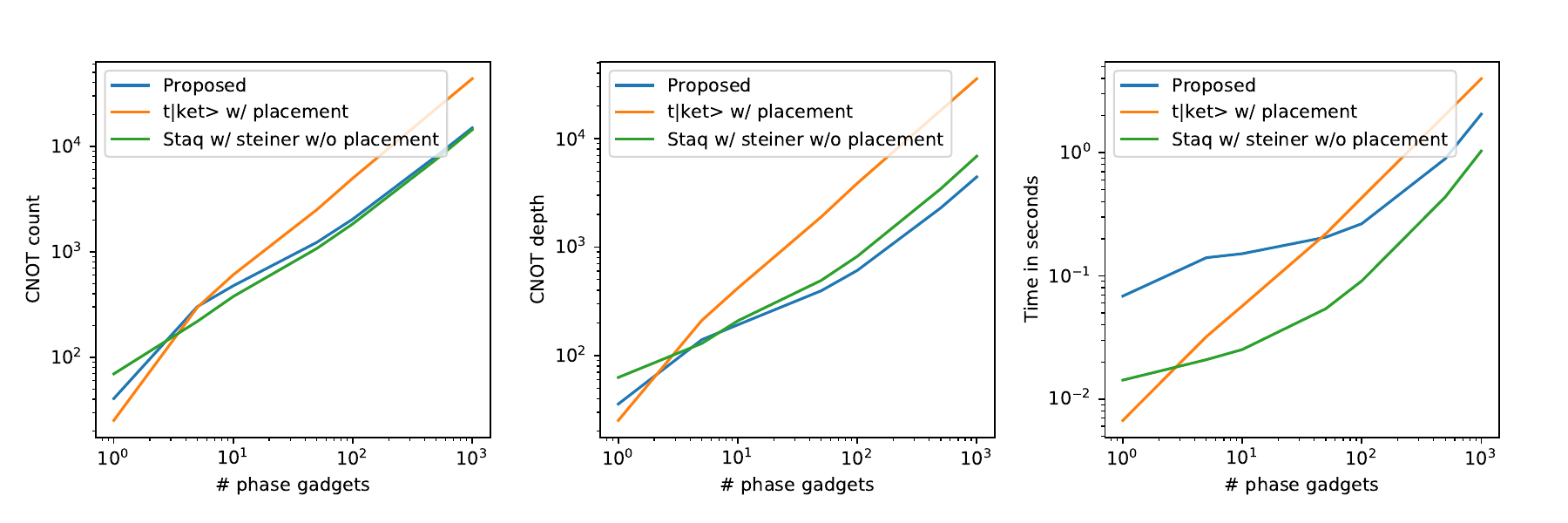}
  \subcaption{20 qubit IBM Singapore}\label{fig:gadgetscale_singapore}
  \end{subfigure}
  \caption{Plots showing the scaling of the \CNOT count, \CNOT depth
    and runtime with respect to the number of phase gadgets on the 16
    qubit Rigetti Aspen architecture and the 20 qubit IBMQ Singapore
    device. 
    The exact data can be found in Table \ref{tbl:real_gadget_scaling} in Appendix \ref{apx:extra_results}.}\label{fig:gadgetscale_real}
  \figureline
\end{figure}

To verify the quality of our algorithm, we generated random phase polynomials 
and synthesised them for two different real quantum computers: 
Rigetti's 16 qubit Aspen device and IBM's 20 qubit Singapore device
\footnote{Qubit-scaling and gadget-scaling results for synthetic architectures can be found in Appendix \ref{apx:extra_results}}. 
We compare the average \CNOT count, \CNOT depth and runtime (in seconds) 
of our proposed algorithm with Staq  \cite{staq} and \tket \cite{TKETPAPERHERE}. 
To the best of our knowledge, \tket and Staq are the only compilers that can 
synthesise and route phase polynomials from an abstract representation\footnote{
  The source code to replicate our results, including the raw experimental data, can be found on \\\url{https://github.com/CQCL/architecture-aware-phasepoly-synth}
}. 

For each architecture, we randomly generated phase
polynomials until we had 20 distinct ones with $1$, $5$, $10$, $50$, $100$, $500$, and $1000$ phase
gadgets in each.  The phase gadgets were sampled uniformly across the
parameter space.  Figure \ref{fig:gadgetscale_real} shows how each
algorithm scales with respect to the number of phase gadgets on the
two quantum computer architectures.  Each point in the chart is the
average of the 20 phase polynomials of that size.

We used pytket version 0.4.3\footnote{This pytket version will be released for the general public soon}. We described our phase polynomials in terms of \tket's abstract representation for phase gadgets (PauliExpBox) which \tket synthesises and then routes using swaps \cite{TKETPAPERHERE}. While routing, we allowed \tket to also find an optimal qubit placement.

For Staq, we used version 1.0. We chose to use the Steiner tree option 
because this results in a much lower \CNOT count and depth. 
Unfortunately, we were unable to use this option in combination with 
optimal qubit placement because this took too long for large phase polynomials ($\geq$ 50 phase gadgets)
\footnote{Staq results with placement for small phase polynomials can be found in Figure \ref{fig:gadgetscale_staq} of Appendix \ref{apx:extra_results}}. 

Note that both \tket and Staq are implemented in C++, while our
algorithm was written in python 3.6, putting it at a significant runtime
disadvantage. All experiments were run on a 2017 MacBook Pro with an
Intel Core i5 2.3 GHz and 8 GB 2133 MHz RAM. We used pytket to
calculate the \CNOT count and \CNOT depth of all circuits (including
Staq).

We observe that for very small phase polynomials (1 phase gadget), 
\tket is the best method, but it does not scale well in \CNOT count and depth 
for larger, more realistic phase polynomials (see Figure \ref{fig:gadgetscale_real}). 
This shows that naive synthesis combined with clever routing is not
competitive with architecture-aware synthesis methods.

Between five and 100 phase gadgets, Staq has the lowest average \CNOT count. 
For larger phase polynomials, Staq's \CNOT count performance is equal to the proposed algorithm.
However, the \CNOT depth is consistently better when synthesised with the proposed algorithm 
for phase polynomials with more than 10 gadgets. 
This means that it is better at parallelising \CNOT gates than Staq.

With respect to runtime, we observe that for phase polynomials with 5-1000 phase gadgets, Staq is the fastest synthesis algorithm. The proposed algorithm is faster at synthesising than \tket for phase polynomials 50-1000 gadgets on both architectures. We do note that both Staq and the proposed algorithm does not scale linearly with respect to the number of phase gadgets, thus it might not be faster than \tket for phase polynomials with more phase gadgets than we have tested.


\section{Conclusion and Future Work}
\label{sec:conc}
In this paper, we introduced one of the first successful algorithms for 
architecture-aware synthesis of phase polynomials. 
We showed that this algorithm performs comparable or better than current state-of-the-art 
compilers for current NISQ devices without compromising the runtime of the algorithm. 

Although our algorithm is very promising, it should still be adjusted to better fit 
the specification of the device that it is synthesising for. 
For example, the choice of placing the qubits affects the size of the synthesised circuit 
because the connectivity graph of a quantum computer is generally not regular. 
Similarly, the current algorithm improves \CNOT depth, 
but it might do so in a way that increases the crosstalk between parallel gates. 

And, lastly, our algorithm can only synthesise phase polynomials. 
This means that circuits containing rotations over X and Y 
need to be split into subcircuits to use our algorithm. 
It will be much more beneficial if our algorithm can be extended to also 
synthesise the generalised version of phase gadgets, called \emph{Pauli exponentials}.

\section*{Acknowledgements}
The authors would like to thank Alex Cowtan, John van de Wetering, 
and Nicolas Heurtel for helpful discussions.
\FloatBarrier
\small
\bibliography{all,arianne}

\newpage
\normalsize
\appendix
\FloatBarrier

\section{Subfunctions for the proposed synthesis algorithm}\label{apx:sub_alg}
The subfunctions that we used in the pseudocode for our algorithm
(Figure \ref{fig:main_alg}) are listed below.

\begin{algorithmic}
    \Function{ReduceColumns}{$Columns$}
    \ForAll {$c \in Columns$}
    \If {$|\{q \in P.\mathrm{rows} \text{ where }  P_{q,c} = 1\}| = 1$}
    \State $Qubit \gets \argmax_{q \in P.\mathrm{rows}} P_{q,c}$
    \State $Circuit$.AddGate($R_Z(ZPhases[c]$, $Qubit$))
    \State $Columns \gets Columns \setminus \{c\}$ 
    \EndIf
    \EndFor
    \State \Return $Columns$
    \EndFunction
\end{algorithmic}

\begin{algorithmic}
    \Function{Place\CNOT}{$Control$, $Target$}
    \State $Circuit$.AddGate(\CNOT($Control$, $Target$))
    \State $P[Control] \gets P[Control]+P[Target]$ 
    \EndFunction
\end{algorithmic}

\begin{algorithmic}
    \Function{SplitColsOnRow}{$Columns$, $Row$}
    \State $Cols0 \gets \{c \in Columns \text{ where }  P_{Row, c} = 0\}$
    \State $Cols1 \gets \{c \in Columns \text{ where } P_{Row, c} = 1\}$
    \State \Return $Cols0$, $Cols1$
    \EndFunction
\end{algorithmic}



\section{Example synthesis}\label{apx:example}
To get a better idea of the inner workings of the algorithm, we synthesise the following phase polynomial:
\begin{align*}
f(\xx) & = \alpha_1(x_2 \oplus x_3) + \alpha_2(x_1) + \alpha_3(x_1 \oplus x_4) + \alpha_4(x_1 \oplus x_2 \oplus x_4) + \alpha_5(x_1 \oplus x_2) + \alpha_6(x_1 \oplus x_2 \oplus x_3) \\
A & = I
\end{align*}
Note that this is the parameterised version of the example phase polynomial given by Amy et al.\cite{Amy_2018}. The connectivity graph we use for synthesis is a simple line architecture: $G : x_1 \Leftrightarrow x_2 \Leftrightarrow x_3 \Leftrightarrow x_4$.

This phase polynomial corresponds to the following ZX-diagram $C$ and matrix representation $P$. 
\[ C \quad = \quad
{\scalebox{0.65}{\inltf{SynthesisExample1}}} \quad \sim \quad  
\begin{pmatrix}
  0 & 1 & 1 & 1 & 1 & 1 \\
  1 & 0 & 0 & 1 & 1 & 1 \\
  1 & 0 & 0 & 0 & 0 & 1 \\
  0 & 0 & 1 & 1 & 0 & 0
\end{pmatrix} \quad = \quad P
\]
Note that the matrix $P$ has a column for each phase gadget in the diagram and each row has a 1 iff the corresponding qubit is participating in the corresponding phase gadget (i.e. it has a green spider). We have added a red vertical line to the ZX-diagram to represent the \emph{frontier}. This indicates the progress of our synthesis. The diagram on the left of the frontier has been synthesised, the diagram on the right of the frontier contains the phase polynomial to be synthesised.
Additionally, while synthesising, we will rewrite the diagram $C$ by adding gates to the frontier without changing the semantics of $C$.

\paragraph{Preprocessing.}
The first step in the algorithm is to check if any columns can be removed from the matrix. This is possible if the column contains exactly a single entry with the value 1. If this is the case, the phase gadget is only acting on a single qubit and it is equivalent to a Z phase gate which we can move to the other side of the frontier.
\begin{align*}
  C & = 
  {\scalebox{0.65}{\inltf{SynthesisExample1}}} 
  = {\scalebox{0.65}{\inltf{SynthesisExample2}}} \\
 & = {\scalebox{0.65}{\inltf{SynthesisExample3}}}
\end{align*}
We describe this process as \emph{placing a phase gate}.

Once the phase gate $R_Z(\alpha_2)$ is placed on qubit $x_1$, we have a phase gadget less, so we can remove the corresponding column from the matrix $P$.
\[ C \quad = \quad
{\scalebox{0.65}{\inltf{SynthesisExample3}}} \quad \sim \quad  
\begin{pmatrix}
  0 & 1 & 1 & 1 & 1 \\
  1 & 0 & 1 & 1 & 1 \\
  1 & 0 & 0 & 0 & 1 \\
  0 & 1 & 1 & 0 & 0
\end{pmatrix}
\]

\paragraph{Main recursion.}

Now we can start the main recursion loop.
We start with the base recursion step and 
calculate all non-cutting vertices of our graph $G$, which are $\{x_1, x_4\}$. 
We pick the row in $P$ with either most ones or most zeroes, which is $x_1$. 
We split the row in to columns with zeroes $P^0 \sim \{\alpha_1\}$, 
and columns with ones $P^1 \sim \{\alpha_3, \alpha_4, \alpha_5, \alpha_6\}$.
We recurse using the base recursion step on $P^0$ and the ones recursion step on $P^1$. 


In the base recursion step on $P^0$, 
we have subgraph $G: x_2 \Leftrightarrow x_3 \Leftrightarrow x_4$, 
with non-cutting vertices $\{x_2, x_4\}$. 
We pick $x_2$ arbitrarily and split the matrix once more 
into $P^{0,0} \sim \emptyset$ and $P^{0,1} \sim \{\alpha_1\}$. 
This time, there are no columns with zeroes, so the base recursion step is trivial. 
Then, in the ones recursion step on $P^{0,1}$, 
we pick a neighbour of $x_2$ with the most ones, 
this is $x_3$, and we place a \CNOT gate, $C_{x_2,x_3}$, in front of the frontier. 
To keep the diagram equivalent to the previous diagrams, 
we add a second \CNOT gate, $C_{x_2,x_3}$, after the frontier and commute it through the phase gadgets. 
By commuting the second \CNOT gate through the gadgets, 
each control qubit will participate in the phase gadget iff either the control or the target qubit (exclusive) was participating before commuting the \CNOT through, 
see Section \ref{sec:related} for a detailed explanation. 
This is the same as adding the target row to the control row in the matrix $P$ (modulo 2). 
Observe that this also changes the columns in $P^1$.
\[ C =
  {\scalebox{0.65}{\inltf{SynthesisExample4}}} 
  = {\scalebox{0.65}{\inltf{SynthesisExample5}}} \quad \sim \quad
  \begin{pmatrix}
    0 & 1 & 1 & 1 & 1 \\
    0 & 0 & 1 & 1 & 0 \\
    1 & 0 & 0 & 0 & 1 \\
    0 & 1 & 1 & 0 & 0
  \end{pmatrix}
  \]
As a result, we can place a phase gate, $R_Z(\alpha_1)$, corresponding to $\alpha_1$ on qubit $x_3$.
  \[C = 
  {\scalebox{0.65}{\inltf{SynthesisExample7}}} \quad \sim \quad
  \begin{pmatrix}
    1 & 1 & 1 & 1 \\
    0 & 1 & 1 & 0 \\
    0 & 0 & 0 & 1 \\
    1 & 1 & 0 & 0
  \end{pmatrix}
  \]
Note that placing the phase gate causes that $P^0 = \emptyset$ so splitting the row $X_2$ 
and recursing on $P^{0,0} \sim \emptyset$ and $P^{0,1} \sim \emptyset$ is trivial.

Now we are finished with the base recursion step on $P^0$ and 
continue with the ones recursion step on $P^1$ and the full graph $G$. 
We had chosen $x_1$ earlier, now we pick a neighbour, $x_2$, and place the \CNOT gate, $C_{x_1, x_2}$. 
This allows us to place a phase gate, $R_Z(\alpha_5)$, on qubit $x_2$.
\[C =
  {\scalebox{0.65}{\inltf{SynthesisExample9}}} 
  = {\scalebox{0.65}{\inltf{SynthesisExample11}}} \sim
  \begin{pmatrix}
    1 & 0 & 1 \\
    0 & 1 & 0 \\
    0 & 0 & 1 \\
    1 & 1 & 0
  \end{pmatrix}
\]

Again, we split row $x_1$ into columns with zeroes $P^{1,0} \sim \{\alpha_4\}$ and 
with ones $P^{1,1} \sim \{\alpha_3, \alpha_6\}$.
We use the base recursion step on $P^{1,0}$ with the subgraph $G: x_2 \Leftrightarrow x_3 \Leftrightarrow x_4$ and we use the ones recursion step on $P^{1,1}$


The  subgraph $G: x_2 \Leftrightarrow x_3 \Leftrightarrow x_4$ has non-cutting vertices $x_2$ and $x_4$. 
We pick $x_4$ arbitrarily and split the row into $P^{1,0, 0} \sim \emptyset$ and $P^{1, 0,1} \sim \{\alpha_4\}$. 
The base recursion step on $P^{1,0,0}$ is trivial.
In the ones recursion step, we pick neighbour $x_3$ and place two \CNOT gates, $C_{x_3,x_4}$, and $C_{x_4, x_3}$, because $x_3$ only has zeroes in $P^{1,0,1}$.

\[C = 
  {\scalebox{0.65}{\inltf{SynthesisExample13}}} \quad \sim \quad
  \begin{pmatrix}
    1 & 0 & 1 \\
    0 & 1 & 0 \\
    1 & 1 & 1 \\
    1 & 1 & 0 \\
  \end{pmatrix}\]
  \[
 C = {\scalebox{0.65}{\inltf{SynthesisExample14}}} \quad \sim \quad
  \begin{pmatrix}
    1 & 0 & 1 \\
    0 & 1 & 0 \\
    1 & 1 & 1 \\
    0 & 0 & 1 \\
  \end{pmatrix}
\]
Now we split $P^{1,0,1}$ on row $x_4$ into $P^{1,0,1,0} \sim \{\alpha_4\}$ 
and $P^{1,0,1,1} \sim \emptyset$ and recurse as before, note that the latter case in trivial.

In the base recursion step on $P^{1,0,1,0}$, we are left with the subgraph $G: x_2 \Leftrightarrow x_3$. 
We pick row $x_2$ arbitrarily and split it into $P^{1,0,1,0,0} \sim \emptyset$ 
and $P^{1,0,1,0,1} \sim \{\alpha_4\}$. 
The base recursion step on $P^{1,0,1,0,0}$ is trivial and 
in the ones recursion step, we pick neighbour $x_3$. 
Hence we can place a \CNOT gate, $C_{x_2, x_3}$, and a phase gate, $R_Z(\alpha_4)$ on qubit $x_3$.
\[C =
  {\scalebox{0.65}{\inltf{SynthesisExample18}}} \quad \sim \quad
  \begin{pmatrix}
    1 & 1 \\
    1 & 1 \\
    1 & 1 \\
    0 & 1 \\
  \end{pmatrix}
\]

This finishes the recursion on $P^{1,0}$ and we can continue with the ones recursion step on $P^{1,1} \sim \{\alpha_3, \alpha_6\}$.
Once more, we are back at the original graph $G: x_1 \Leftrightarrow x_2 \Leftrightarrow x_3 \Leftrightarrow x_4$. 
We previously picked row $x_1$ and so we now pick neighbour $x_2$.
We place a \CNOT gate, $C_{x_1, x_2}$, and split on row $x_1$ into $P^{1,1,0} \sim \{\alpha_3, \alpha_6\}$, and $P^{1,1,1} \sim \emptyset$.

\[C =
  {\scalebox{0.65}{\inltf{SynthesisExample20}}} \quad \sim \quad
  \begin{pmatrix}
    0 & 0 \\
    1 & 1 \\
    1 & 1 \\
    0 & 1 \\
  \end{pmatrix} 
\]

In the base recursion step on $P^{1,1,0}$, we pick row $x_2$ and split $P^{1,1,0}$ into
$P^{1,1,0,0} \sim \emptyset$, and $P^{1,1,0,1} \sim \{\alpha_3, \alpha_6\}$. The base recursion step on $P^{1,1,0,0}$ is trivial.

In the ones recursion step on $P^{1,1,0,1}$, we pick neighbour $x_3$, and place a \CNOT  gate, $C_{x_2, x_3}$, and a phase gate, $R_Z(\alpha_3)$, on qubit $x_3$.
  
\[C =
  {\scalebox{0.65}{\inltf{SynthesisExample24}}} \quad \sim \quad
  \begin{pmatrix}
    0 \\
    0 \\
    1 \\
    1 \\
  \end{pmatrix}
\]
We split $P^{1,1,0,1}$ on row $x_2$, resulting in $P^{1,1,0,1,0} \sim \{\alpha_6\}$, and $P^{1,1,0,1,1} \sim \emptyset$ and we recurse as before.

In the base recursion on $P^{1,1,0,1,0}$,
we are left with subgraph $G: x_3 \Leftrightarrow x_4$. 
We pick $x_3$ and split on it resulting in $P^{1,1,0,1,0,0} \sim \emptyset$ and $P^{1,1,0,1,0,1} \sim \{\alpha_6\}$. The base recursion step is trivial.

Finally, in the ones recursion step on $P^{1,1,0,1,0,1}$,
we pick neighbour $x_4$ and place a \CNOT gate, $C_{x_3, x_4}$ and a phase gate, $R_Z(\alpha_6)$, on qubit $x_4$.
  
\[C =
  {\scalebox{0.65}{\inltf{SynthesisExample28}}} \quad \sim \quad
  \begin{pmatrix}
    \\
    \\
    \\
    \\
  \end{pmatrix}  
\]
Now we have synthesised every phase gadget in the support of $f$.

\paragraph{Post-processing.}
What remains is synthesising the basis transform $A = I$.
At the frontier, the basis transform of the qubits is equal to the matrix $P'$,
\[ P' =
  \begin{pmatrix}
    1&0&1&1\\
    0&1&0&1\\
    0&0&0&1\\
    0&0&1&0\\
  \end{pmatrix}
  \]
as can be seen in the parity annotation of each qubit of the final circuit.
This transform needs to be undone before the basis transform $A$ can be applied.

As explained at the end of Section \ref{sec:methods}, this is transformation is undone by the trailing \CNOTs on the right of the frontier.
I.e. the \CNOTs on the right of the frontier apply the basis transform $P'^{-1}$.
Although these \CNOTs are already mapped, they could be optimised using an architecture-aware \CNOT circuit synthesis technique, such as Steiner-Gauss. In case the matrix $A \neq I$, we can calculate the full transformation $A'$ by undoing the existing linear transformation and then applying the desired transformation: $A' = A \cdot P'^{-1}$.

\FloatBarrier
\section{Additional results}\label{apx:extra_results}
This appendix contains additional figures and tables to show the performance of the proposed algorithm with respect to the existing algorithms.

To show the scaling of our algorithm with respect the number of qubits, the number of phase gadgets and the density of the device connectivity graph, we have run several experiments, generating 20 random phase polynomials per experimental setting. 
Since Staq only supports a small selection of quantum computer architectures, we compare the proposed algorithm against an in-house implementation of Steiner-GraySynth for all synthetic architectures.

\begin{figure}[t]
  \centering
  \begin{subfigure}[b]{\textwidth}
  \includegraphics[width=\textwidth]{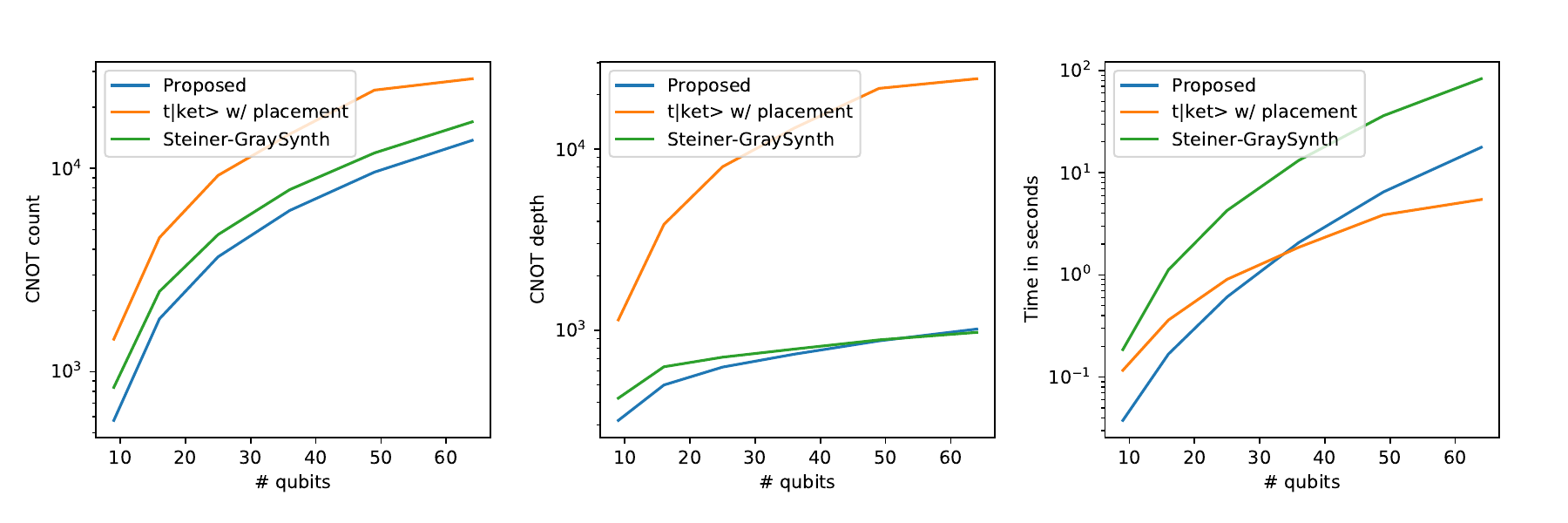}
  \subcaption{line}\label{fig:qubitscale_line}
  \end{subfigure}
  \begin{subfigure}[b]{\textwidth}
  \includegraphics[width=\textwidth]{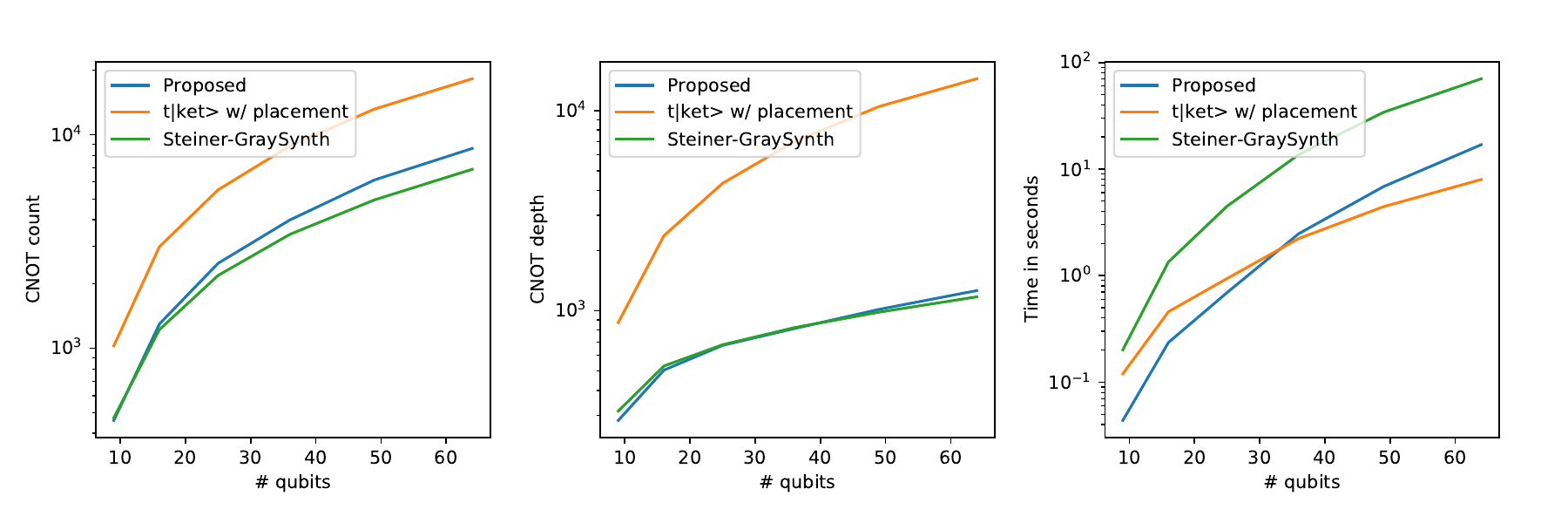}
  \subcaption{square grid}\label{fig:qubitscale_square}
  \end{subfigure}
  \begin{subfigure}[b]{\textwidth}
  \includegraphics[width=\textwidth]{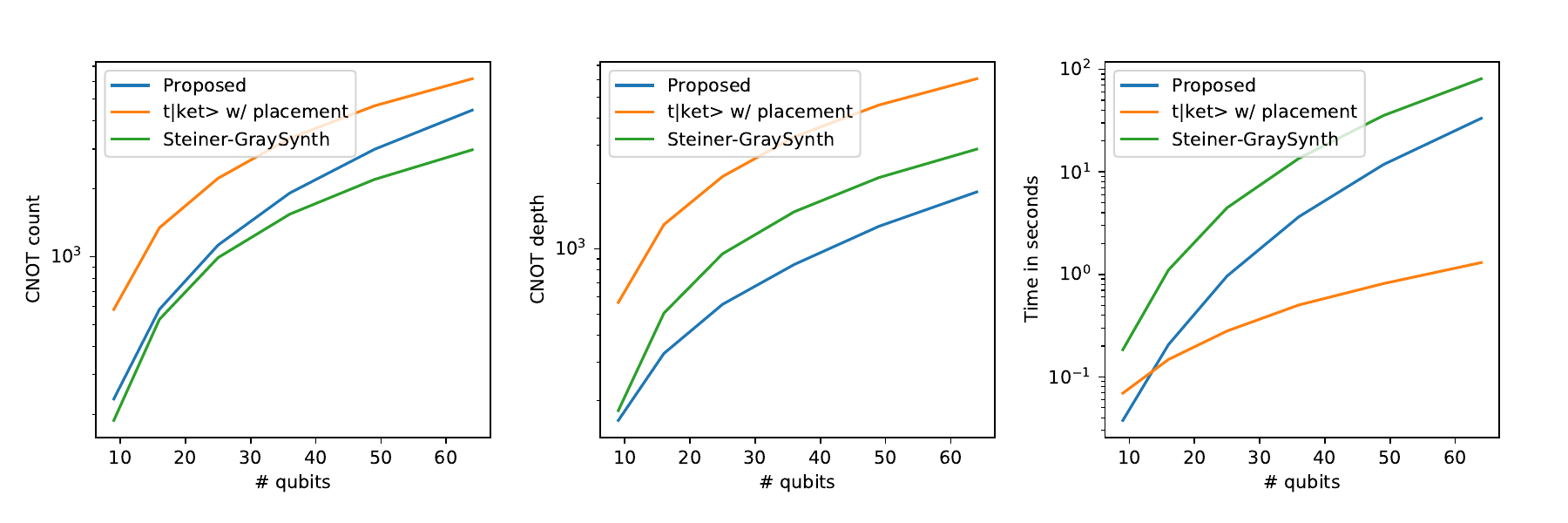}
  \subcaption{fully connected}\label{fig:qubitscale_fully}
  \end{subfigure}
  \caption{The influence of the number of qubits on the \CNOT count, \CNOT depth and runtime for architectures with different regular structures: line, square grid and fully connected.
  The exact data can be found in Table \ref{tbl:fake_qubit_scaling}.
  }\label{fig:qubitscale}
\end{figure}

\begin{figure}[t]
  \centering
  \begin{subfigure}[b]{\textwidth}
  \includegraphics[width=\textwidth]{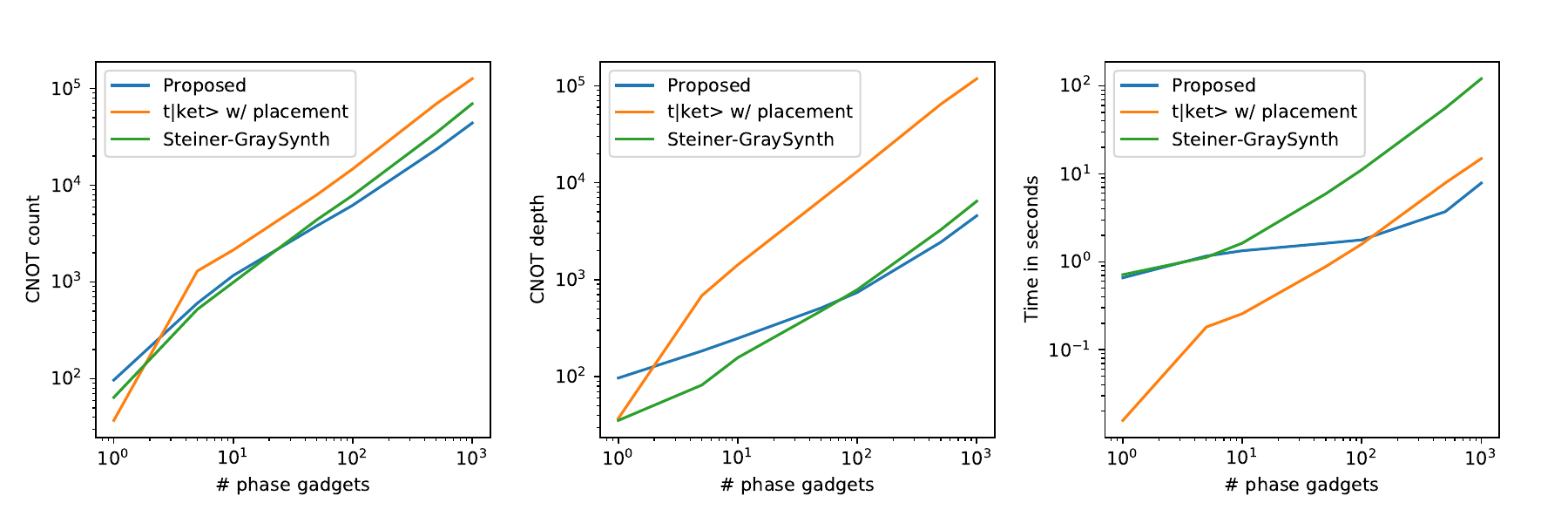}
  \subcaption{line}\label{fig:gadgetscale_line}
  \end{subfigure}
  \begin{subfigure}[b]{\textwidth}
  \includegraphics[width=\textwidth]{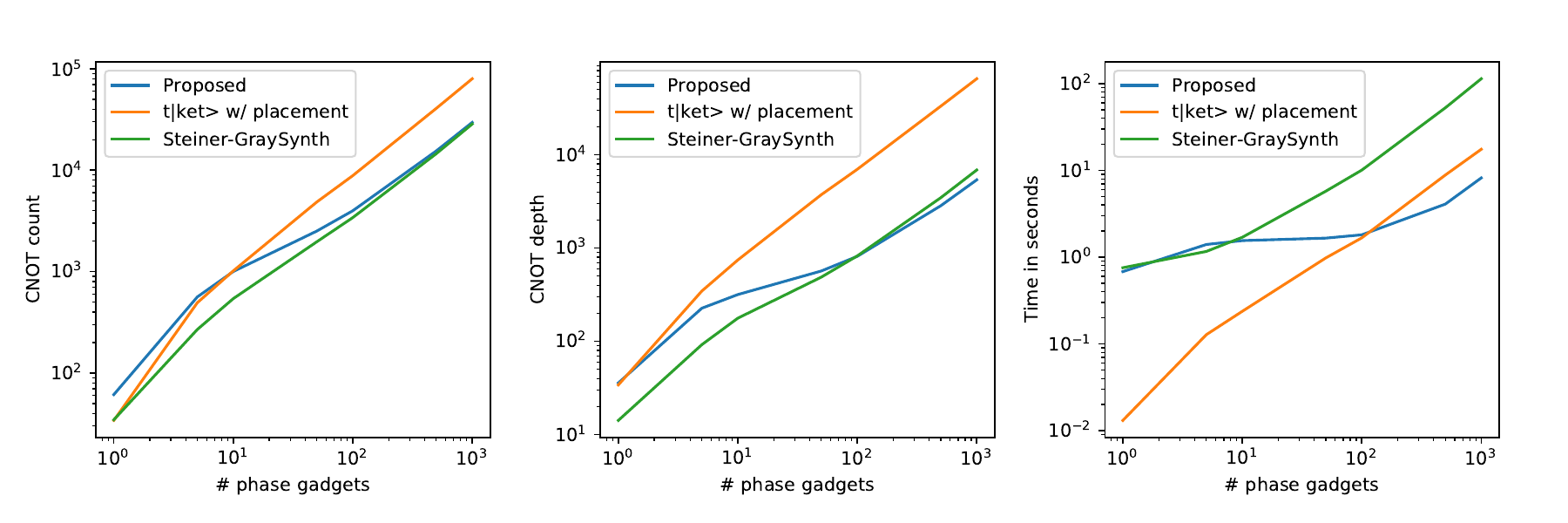}
  \subcaption{square grid}\label{fig:gadgetscale_square}
  \end{subfigure}
  \begin{subfigure}[b]{\textwidth}
  \includegraphics[width=\textwidth]{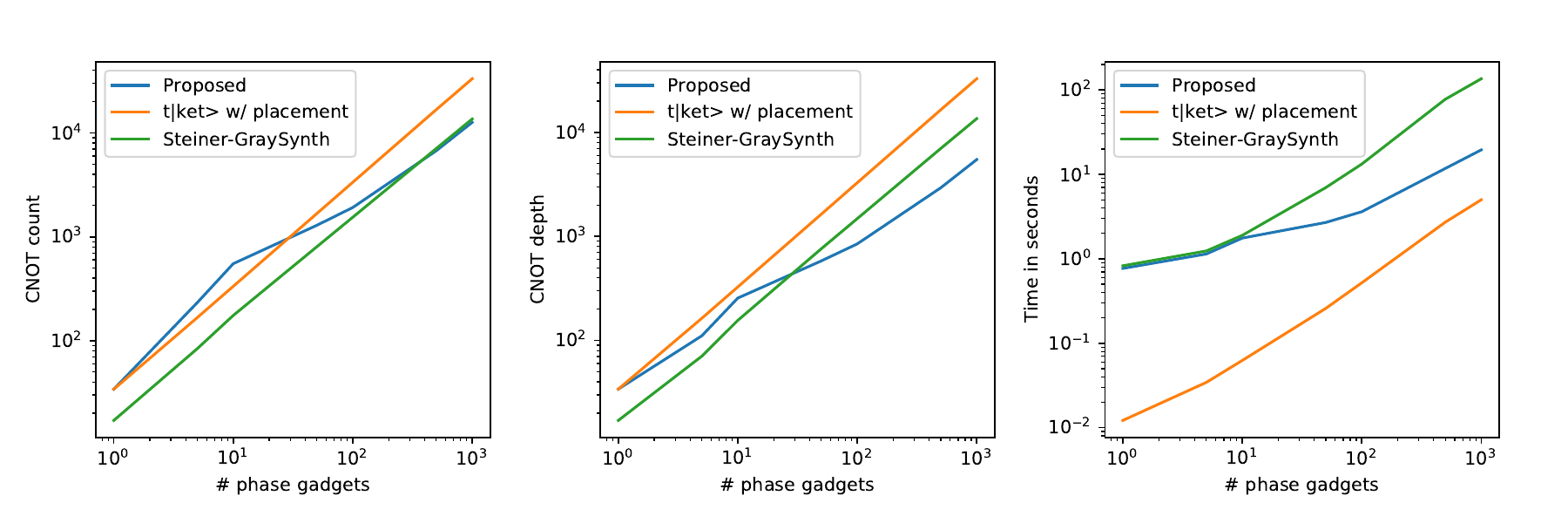}
  \subcaption{fully connected}\label{fig:gadgetscale_fully}
  \end{subfigure}
  \caption{Plots showing the scaling of the \CNOT count, \CNOT depth and runtime with respect to the number of phase gadgets on a 36 qubit line, square grid, and unconstrained architecture.
  The exact data can be found in Table \ref{tbl:fake_gadget_scaling}.}\label{fig:gadgetscale_fake}
\end{figure}

\begin{figure}[t]
  \centering
  \begin{subfigure}[b]{\textwidth}
  \includegraphics[width=\textwidth]{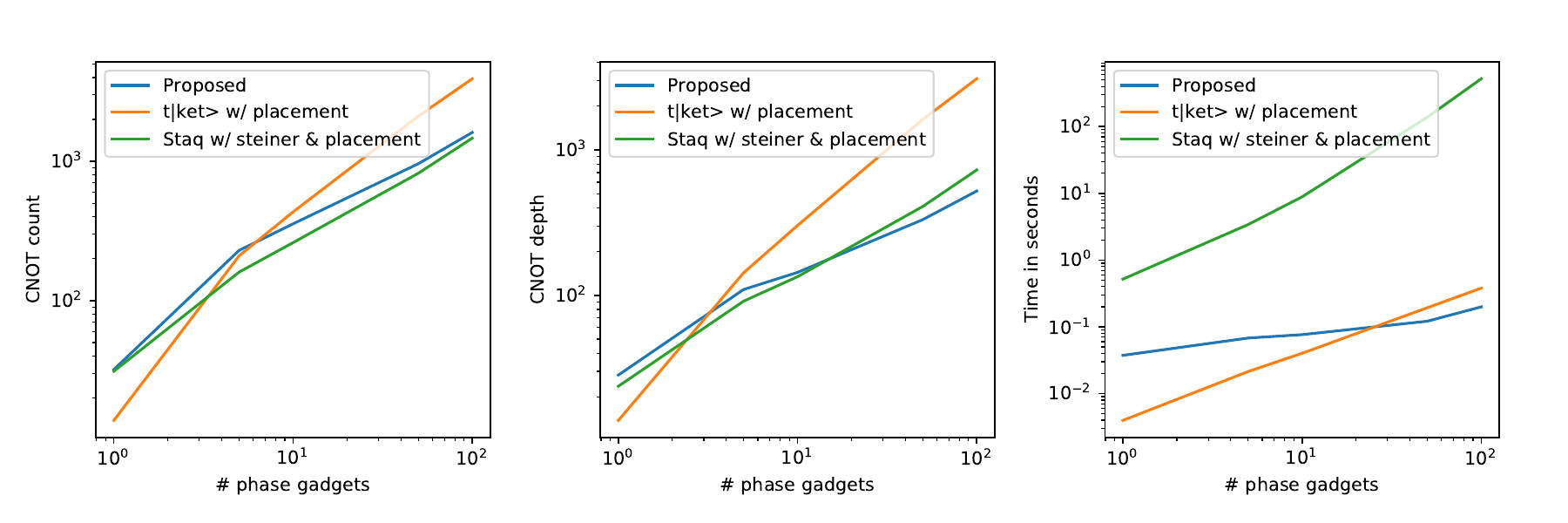}
  \subcaption{16 qubit Rigetti Aspen}\label{fig:gadgetscale_staq_aspen}
  \end{subfigure}
  \begin{subfigure}[b]{\textwidth}
  \includegraphics[width=\textwidth]{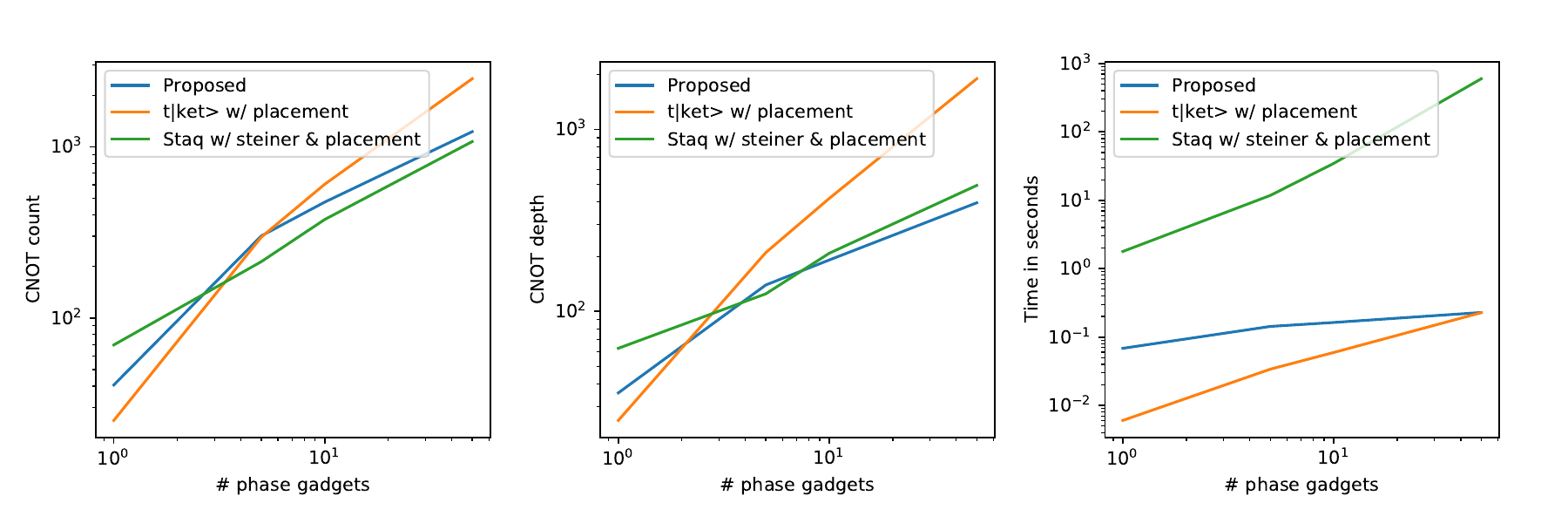}
  \subcaption{20 qubit IBM Singapore}\label{fig:gadgetscale_staq_singapore}
  \end{subfigure}
  \caption{Plots showing the scaling of the \CNOT count, \CNOT depth and runtime with respect to the number of phase gadgets on the 16 qubit Rigetti Aspen architecture and the 20 qubit IBMQ Singapore device.
  The exact data can be found in Table \ref{tbl:real_gadget_scaling_mapping}.}\label{fig:gadgetscale_staq}
\end{figure}

Figure \ref{fig:qubitscale} shows how our algorithm and the two baselines perform on a line, square and fully connected connectivity of various sizes given a phase polynomial with 100 phase gadgets.
Similarly, Figure \ref{fig:gadgetscale_fake} shows how our algorithm and the two baselines perform on phase polynomials of various sizes given a 36 qubit line, square and unconstrained connectivity graph.
In Figure \ref{fig:gadgetscale_staq}, we show that, if Staq is used with qubit placement optimisation, it can synthesise slightly smaller circuits than without qubit placement. However, this comes at an extreme runtime cost. The runtime of this option was long enough that it was not feasible for to run experiments with more than 50 and 100 gadgets (IBMQ Singapore and Rigetti Aspen, respectively) because Staq would take more than two hours to synthesise a single circuit with 500 gadgets on Rigetti Aspen.

Lastly, the exact data that was visualised in each figure, Figure \ref{fig:gadgetscale_real}, \ref{fig:qubitscale}, \ref{fig:gadgetscale_fake}, and \ref{fig:gadgetscale_staq}, is given in Table \ref{tbl:real_gadget_scaling}, \ref{tbl:fake_qubit_scaling}, \ref{tbl:fake_gadget_scaling}, and \ref{tbl:real_gadget_scaling_mapping}, respectively.

\FloatBarrier

\begin{table}
  \scalebox{0.85}{
\begin{subtable}[t]{\textwidth}
\begin{tabular}{l|ccc|ccc|ccc}%
   & & \textbf{\tket} & & & \bfseries Staq & & & \bfseries Proposed \\
   \textbf{$\# R_Z$} & count & depth & time & count & depth & time & count & depth & time\\\hline\hline
  \begin{filecontents*}{data/aggr_rigetti_16q_aspen_gadget_scaling.csv}
gadgets,proposedcount,staqcount,tketcount,proposeddepth,staqdepth,tketdepth,proposedtime,staqtime,tkettime
1,32.00,31.10,13.80,28.30,23.70,13.80,0.035,0.017,0.004
5,229.30,169.85,209.55,109.70,94.70,142.95,0.066,0.018,0.020
10,355.80,264.80,433.25,143.95,136.45,302.65,0.073,0.021,0.038
50,961.05,820.80,2109.35,331.90,409.00,1622.45,0.110,0.044,0.164
100,1611.90,1466.85,3917.60,522.05,728.80,3090.60,0.151,0.074,0.306
500,6081.30,5928.60,17416.70,2082.25,3293.90,14274.65,0.611,0.345,1.506
1000,11238.30,11043.40,32357.05,4018.20,6500.05,26896.25,1.431,0.807,2.851
\end{filecontents*}
\csvreader[head to column names, late after line={\\\hline}]{data/aggr_rigetti_16q_aspen_gadget_scaling.csv}{}
  {\gadgets & \tketcount & \tketdepth & \tkettime s & \staqcount & \staqdepth & \staqtime s & \proposedcount & \proposeddepth & \proposedtime s}
\end{tabular}
\subcaption{Rigetti 16Q Aspen}\label{tbl:real_gadget_scaling_aspen}
\end{subtable}
}
\\
\scalebox{0.85}{

\begin{subtable}[t]{\textwidth}
\begin{tabular}{l|ccc|ccc|ccc}%
  & & \textbf{\tket} & & & \bfseries Staq & & & \bfseries Proposed \\
  \textbf{$\# R_Z$} & count & depth & time & count & depth & time & count & depth & time \\\hline\hline
 \begin{filecontents*}{data/aggr_ibmq_singapore_gadget_scaling.csv}
gadgets,proposedcount,staqcount,tketcount,proposeddepth,staqdepth,tketdepth,proposedtime,staqtime,tkettime
1,40.60,69.60,25.20,35.70,62.75,25.20,0.068,0.014,0.007
5,301.60,218.60,296.70,139.75,129.15,210.60,0.140,0.021,0.032
10,475.50,376.50,604.95,191.75,208.60,417.90,0.151,0.025,0.057
50,1226.00,1073.25,2499.30,395.20,492.40,1897.70,0.206,0.054,0.219
100,2035.10,1834.75,4969.60,607.95,819.65,3863.90,0.265,0.091,0.431
500,8054.35,7498.85,22710.00,2293.45,3440.55,18205.90,0.893,0.437,2.033
1000,14908.55,14309.70,43538.10,4422.70,6867.05,35533.30,2.054,1.031,3.983
\end{filecontents*}
\csvreader[head to column names, late after line={\\\hline}]{data/aggr_ibmq_singapore_gadget_scaling.csv}{}
 {\gadgets & \tketcount & \tketdepth & \tkettime s & \staqcount & \staqdepth & \staqtime s & \proposedcount & \proposeddepth & \proposedtime s }
\end{tabular}
\subcaption{IBMQ Singapore}\label{tbl:real_gadget_scaling_singapore}
\end{subtable}
}
\caption{The average number of CNOT, CNOT depth and runtime for 20 circuits for synthesising phase polynomials with various sizes using \tket, Staq (without qubit placement) and our proposed algorithm on Rigetti Aspen (Table \ref{tbl:real_gadget_scaling_aspen}) and IBMQ Singapore (Table \ref{tbl:real_gadget_scaling_singapore}). This data was visualised in Figure \ref{fig:gadgetscale_real}.}\label{tbl:real_gadget_scaling}
  
\end{table}

\begin{table}  \scalebox{0.85}{
  \begin{subtable}[t]{0.9\textwidth}
  \begin{tabular}{l|ccc|ccc|ccc}%
    & & \textbf{\tket} & & & \bfseries Nash & & & \bfseries Proposed \\
    \textbf{Qubits} & count & depth & time & count & depth & time & count & depth & time \\\hline\hline
   \begin{filecontents*}{data/aggr_line_qubit_scaling.csv}
qubits,nashcount,proposedcount,tketcount,nashdepth,proposeddepth,tketdepth,nashtime,proposedtime,tkettime
9,836.45,575.40,1444.30,422.75,318.65,1142.75,0.186,0.038,0.117
16,2480.10,1818.75,4565.50,629.95,499.85,3846.85,1.121,0.168,0.362
25,4724.60,3673.35,9240.80,711.55,627.50,8004.30,4.269,0.609,0.905
36,7866.50,6211.55,14761.20,788.75,739.30,13074.70,13.198,2.072,1.863
49,11920.00,9592.70,24291.45,886.60,875.40,21651.35,36.099,6.484,3.867
64,16960.45,13750.00,27632.10,975.30,1017.30,24473.65,82.994,17.706,5.462
\end{filecontents*}
\csvreader[head to column names, late after line={\\\hline}]{data/aggr_line_qubit_scaling.csv}{}
   {\qubits & \tketcount & \tketdepth & \tkettime s & \nashcount & \nashdepth & \nashtime s & \proposedcount & \proposeddepth & \proposedtime s }
  \end{tabular}
  \subcaption{Line}\label{tbl:fake_qubit_scaling_line}
  \end{subtable}

  }
  \\
  \scalebox{0.85}{
  \begin{subtable}[t]{0.9\textwidth}
  \begin{tabular}{l|ccc|ccc|ccc}%
     & & \textbf{\tket} & & & \bfseries Nash & & & \bfseries Proposed \\
     \textbf{Qubits} & count & depth & time & count & depth & time & count & depth & time \\\hline\hline
    \begin{filecontents*}{data/aggr_square_qubit_scaling.csv}
qubits,nashcount,proposedcount,tketcount,nashdepth,proposeddepth,tketdepth,nashtime,proposedtime,tkettime
9,472.15,459.35,1025.65,316.10,283.85,874.75,0.201,0.044,0.120
16,1222.50,1299.00,2986.20,530.50,505.95,2372.30,1.342,0.236,0.458
25,2191.35,2497.65,5514.35,677.30,672.35,4327.40,4.479,0.693,0.939
36,3409.85,3982.25,8842.65,824.35,815.60,6993.00,13.679,2.475,2.227
49,4948.05,6135.10,13171.00,984.55,1016.65,10457.15,34.106,6.843,4.432
64,6881.55,8615.05,18259.70,1174.30,1261.85,14393.05,70.358,16.918,7.977
\end{filecontents*}
\csvreader[head to column names, late after line={\\\hline}]{data/aggr_square_qubit_scaling.csv}{}
    {\qubits & \tketcount & \tketdepth & \tkettime s & \nashcount & \nashdepth & \nashtime s & \proposedcount & \proposeddepth & \proposedtime s }
  \end{tabular}
  \subcaption{Square}\label{tbl:fake_qubit_scaling_square}
  \end{subtable}
  
  }
  \\
  \scalebox{0.85}{
  \begin{subtable}[t]{\textwidth}
  \begin{tabular}{l|ccc|ccc|ccc}%
    & & \textbf{\tket} & & & \bfseries Nash & & & \bfseries Proposed \\
    \textbf{Qubits} & count & depth & time & count & depth & time & count & depth & time \\\hline\hline
   \begin{filecontents*}{data/aggr_fully_connected_qubit_scaling.csv}
qubits,nashcount,proposedcount,tketcount,nashdepth,proposeddepth,tketdepth,nashtime,proposedtime,tkettime
9,187.75,233.75,583.30,180.05,162.25,566.90,0.184,0.038,0.069
16,527.05,583.85,1343.40,506.25,330.15,1294.60,1.103,0.206,0.148
25,990.10,1125.55,2227.70,949.85,555.80,2155.15,4.463,0.963,0.280
36,1543.40,1915.45,3351.10,1483.15,847.85,3281.25,13.501,3.642,0.502
49,2200.75,2994.10,4668.50,2131.25,1271.05,4599.60,35.417,11.777,0.813
64,2978.95,4466.60,6155.90,2880.10,1829.50,6076.20,80.856,33.257,1.303
\end{filecontents*}
\csvreader[head to column names, late after line={\\\hline}]{data/aggr_fully_connected_qubit_scaling.csv}{}
   {\qubits & \tketcount & \tketdepth & \tkettime s & \nashcount & \nashdepth & \nashtime s & \proposedcount & \proposeddepth & \proposedtime s }
  \end{tabular}
  \subcaption{Unconstrained}\label{tbl:fake_qubit_scaling_fully}
  \end{subtable}
  }
  \caption{The average number of CNOT, CNOT depth and runtime for 20 circuits for synthesising phase polynomials with 100 phase gadgets using \tket, Nash and our proposed algorithm on synthetic qubit architectures of various sizes connected in a line (Table \ref{tbl:fake_qubit_scaling_line}), square (Table \ref{tbl:fake_qubit_scaling_square}) and fully connected (Table \ref{tbl:fake_qubit_scaling_fully}).
  This data was visualised in Figure \ref{fig:qubitscale}.}\label{tbl:fake_qubit_scaling}
\end{table}

\begin{table}
  \scalebox{0.85}{
  \begin{subtable}[t]{\textwidth}
    \centering
  \begin{tabular}{l|ccc|ccc|ccc}%
    & & \textbf{\tket} & & & \bfseries Nash & & & \bfseries Proposed \\
    \textbf{$\# R_Z$} & count & depth & time & count & depth & time & count & depth & time \\\hline\hline
   \begin{filecontents*}{data/aggr_line_gadget_scaling.csv}
gadgets,nashcount,proposedcount,tketcount,nashdepth,proposeddepth,tketdepth,nashtime,proposedtime,tkettime
1,64.15,96.90,37.10,35.40,96.90,37.10,0.714,0.658,0.016
5,521.10,603.80,1298.70,82.00,183.95,684.85,1.126,1.163,0.181
10,991.85,1167.45,2142.80,157.15,248.40,1424.30,1.631,1.335,0.257
50,4391.70,3800.50,7960.45,477.30,512.50,6700.05,5.943,1.624,0.884
100,7866.50,6211.55,14761.20,788.75,739.30,13074.70,11.137,1.776,1.591
500,34883.95,23425.10,69417.10,3272.30,2435.75,64493.35,55.913,3.718,7.885
1000,69584.50,43934.05,126095.50,6446.50,4564.45,118035.55,120.575,7.853,14.913
\end{filecontents*}
\csvreader[head to column names, late after line={\\\hline}]{data/aggr_line_gadget_scaling.csv}{}
   {\gadgets & \tketcount & \tketdepth & \tkettime s & \nashcount & \nashdepth & \nashtime s & \proposedcount & \proposeddepth & \proposedtime s }
  \end{tabular}
  \subcaption{36 qubit line}\label{tbl:fake_gadget_scaling_line}
  \end{subtable}
  }
  \\
  \scalebox{0.85}{

  \begin{subtable}[t]{\textwidth}
    \centering
  \begin{tabular}{l|ccc|ccc|ccc}%
     & & \textbf{\tket} & & & \bfseries Nash & & & \bfseries Proposed \\
     \textbf{$\# R_Z$} & count & depth & time & count & depth & time & count & depth & time\\\hline\hline
    \begin{filecontents*}{data/aggr_square_gadget_scaling.csv}
gadgets,nashcount,proposedcount,tketcount,nashdepth,proposeddepth,tketdepth,nashtime,proposedtime,tkettime
1,34.60,61.30,34.10,14.20,35.80,34.10,0.755,0.681,0.013
5,268.80,562.95,493.45,92.15,226.65,346.30,1.164,1.400,0.128
10,541.70,1002.75,1017.65,177.15,317.45,745.20,1.695,1.552,0.238
50,1969.35,2512.10,4859.05,488.35,568.20,3737.90,5.762,1.658,0.980
100,3409.85,3982.25,8842.65,824.35,815.60,6993.00,10.080,1.809,1.669
500,14560.45,15492.85,40734.55,3469.65,2844.85,33232.65,52.656,4.087,8.840
1000,28530.30,29621.30,79821.90,6876.90,5390.25,65423.70,113.956,8.196,17.518
\end{filecontents*}
\csvreader[head to column names, late after line={\\\hline}]{data/aggr_square_gadget_scaling.csv}{}
    {\gadgets & \tketcount & \tketdepth & \tkettime s & \nashcount & \nashdepth & \nashtime s & \proposedcount & \proposeddepth & \proposedtime s }
  \end{tabular}
  \subcaption{36 qubit square}\label{tbl:fake_gadget_scaling_square}
  \end{subtable}
  
  }
  \\
  \scalebox{0.85}{
  \begin{subtable}[t]{\textwidth}
  \begin{tabular}{l|ccc|ccc|ccc}%
    & & \textbf{\tket} & & & \bfseries Nash & & & \bfseries Proposed \\
    \textbf{$\# R_Z$} & count & depth & time & count & depth & time & count & depth & time \\\hline\hline
   \begin{filecontents*}{data/aggr_fully_connected_gadget_scaling.csv}
gadgets,nashcount,proposedcount,tketcount,nashdepth,proposeddepth,tketdepth,nashtime,proposedtime,tkettime
1,17.05,34.10,34.10,17.05,34.10,34.10,0.826,0.769,0.012
5,83.80,232.70,166.90,70.65,110.90,163.50,1.239,1.141,0.034
10,174.95,551.40,333.80,156.00,255.60,326.95,1.897,1.759,0.063
50,801.60,1290.25,1673.00,761.30,581.80,1636.20,6.981,2.695,0.259
100,1543.40,1915.45,3351.10,1483.15,847.85,3281.25,13.252,3.611,0.520
500,7081.75,6741.75,16781.80,7006.85,2935.00,16489.55,77.596,11.749,2.715
1000,13642.20,12660.40,33291.10,13550.90,5479.05,32759.35,135.732,19.504,4.999
\end{filecontents*}
\csvreader[head to column names, late after line={\\\hline}]{data/aggr_fully_connected_gadget_scaling.csv}{}
   {\gadgets & \tketcount & \tketdepth & \tkettime s & \nashcount & \nashdepth & \nashtime s & \proposedcount & \proposeddepth & \proposedtime s }
  \end{tabular}
  \subcaption{36 qubit unconstrained}\label{tbl:fake_gadget_scaling_fully}
  \end{subtable}
  }
  \caption{The average number of CNOT, CNOT depth and runtime for 20 circuits for synthesising phase polynomials with various sizes using \tket, Nash and our proposed algorithm on synthetic 36 qubit architectures connected in a line (Table \ref{tbl:fake_gadget_scaling_line}), square (Table \ref{tbl:fake_gadget_scaling_square}) and fully connected (Table \ref{tbl:fake_gadget_scaling_fully}).
  This data was visualised in Figure \ref{fig:gadgetscale_fake}.}\label{tbl:fake_gadget_scaling}
    
\end{table}

\begin{table}
  \scalebox{0.85}{
  \begin{subtable}[t]{\textwidth}
  \begin{tabular}{l|ccc|ccc|ccc}%
     & & \textbf{\tket} & & & \bfseries Staq & & & \bfseries Proposed \\
     \textbf{$\# R_Z$} & count & depth & time & count & depth & time & count & depth & time\\\hline\hline
    \begin{filecontents*}{data/aggr_rigetti_16q_aspen_gadget_scaling_with_mapping.csv}
gadgets,proposedcount,staqcount,tketcount,proposeddepth,staqdepth,tketdepth,proposedtime,staqtime,tkettime
1,32.00,31.10,13.80,28.30,23.70,13.80,0.037,0.518,0.004
5,229.30,160.20,209.55,109.70,91.20,142.95,0.068,3.407,0.021
10,355.80,260.65,433.25,143.95,134.45,302.65,0.076,8.854,0.040
50,961.05,820.80,2109.35,331.90,409.00,1622.45,0.121,138.448,0.193
100,1611.90,1466.85,3917.60,522.05,728.80,3090.60,0.199,521.102,0.380
\end{filecontents*}
\csvreader[head to column names, late after line={\\\hline}]{data/aggr_rigetti_16q_aspen_gadget_scaling_with_mapping.csv}{}
    {\gadgets & \tketcount & \tketdepth & \tkettime s & \staqcount & \staqdepth & \staqtime s & \proposedcount & \proposeddepth & \proposedtime s}
  \end{tabular}
  \subcaption{Rigetti 16Q Aspen}\label{tbl:real_gadget_scaling_mapping_aspen}
  \end{subtable}
  }
  \\
  \scalebox{0.85}{
  \begin{subtable}[t]{\textwidth}
  \begin{tabular}{l|ccc|ccc|ccc}%
    & & \textbf{\tket} & & & \bfseries Staq & & & \bfseries Proposed \\
    \textbf{$\# R_Z$} & count & depth & time & count & depth & time & count & depth & time \\\hline\hline
   \begin{filecontents*}{data/aggr_ibmq_singapore_gadget_scaling_with_mapping.csv}
gadgets,proposedcount,staqcount,tketcount,proposeddepth,staqdepth,tketdepth,proposedtime,staqtime,tkettime
1,40.60,69.60,25.20,35.70,62.75,25.20,0.068,1.784,0.006
5,301.60,213.30,296.70,139.75,124.95,210.60,0.143,11.770,0.034
10,475.50,376.30,604.95,191.75,208.80,417.90,0.163,34.636,0.060
50,1226.00,1073.25,2499.30,395.20,492.40,1897.70,0.229,597.870,0.227
\end{filecontents*}
\csvreader[head to column names, late after line={\\\hline}]{data/aggr_ibmq_singapore_gadget_scaling_with_mapping.csv}{}
   {\gadgets & \tketcount & \tketdepth & \tkettime s & \staqcount & \staqdepth & \staqtime s & \proposedcount & \proposeddepth & \proposedtime s }
  \end{tabular}
  \subcaption{IBMQ Singapore}\label{tbl:real_gadget_scaling_mapping_singapore}
  \end{subtable}
  }
  \caption{The average number of CNOT, CNOT depth and runtime for 20 circuits for synthesising phase polynomials with various sizes using \tket, Staq (with qubit placement) and our proposed algorithm on Rigetti Aspen (Table \ref{tbl:real_gadget_scaling_mapping_aspen}) and IBMQ Singapore (Table \ref{tbl:real_gadget_scaling_mapping_singapore}).
  This data was visualised in Figure \ref{fig:gadgetscale_staq}.}\label{tbl:real_gadget_scaling_mapping}
    
  \end{table}
  
\end{document}

%% file: tikzpreamble.tex
\usepackage{tikzit}  

\newcommand{\inlinetikzfig}[2][1.0]{
  \dimendef\grafheight=255\dimendef\grafvshift=254
  \setbox0 = \hbox{\scalebox{#1}{\input{#2.tikz}}}
  \grafheight=\the\ht0
  \grafvshift=-0.5\grafheight
  \advance\grafvshift by 0.5ex
  \raisebox{\grafvshift}{\input{#2.tikz}}
}

\newcommand{\inltf}[1]{\inlinetikzfig{#1}}

\pgfdeclarelayer{background} 
\pgfdeclarelayer{foreground}
\pgfdeclarelayer{edgelayer}
\pgfdeclarelayer{nodelayer}
\pgfsetlayers{background,edgelayer,nodelayer,main,foreground} 

\tikzstyle{halfsize}=[x=0.5cm, y=0.5cm]
\tikzstyle{normalsize}=[]
\tikzstyle{doublesize}=[]

\tikzstyle{(null)}=[]
\tikzstyle{plain}=[]

\usepackage{tikzzx}
\usepackage{tikzcircuits}
\usepackage[undirected]{tikzquanto} 
\input{general.tikzstyles}
\usetikzlibrary{calc}

\tikzset{every picture/.style={line width=0.75pt}} 
\tikzset{directed edge/.style={postaction={decorate,decoration={markings,mark=at position 0.5 with {\arrow{>}}}}}}

%% file: general.tikzstyles
\tikzstyle{double arrow}=[<->, >=triangle 45, line width=1pt]
\tikzstyle{single arrow}=[->, >=triangle 45, line width=1pt]

%% file: preamble.tex
\usepackage[T1]{fontenc}
\usepackage{fixltx2e}
\usepackage{microtype}
\usepackage{xspace}
\usepackage{enumitem}

\newcommand{\figureline}{\rule{\textwidth}{0.5pt}}

\makeatletter
\newcommand\etc{etc\@ifnextchar.{}{.\@}\xspace}

\makeatother

\usepackage{graphicx}

\newcommand{\inlinegraphic}[2]{
  \dimendef\grafheight=255\dimendef\grafvshift=254
  \grafheight=#1
  \grafvshift=-0.5\grafheight
  \advance\grafvshift by 0.5ex
  \raisebox{\grafvshift}{\includegraphics[height=\grafheight]{images/#2}\xspace}
}

\newcommand{\ninlinegraphic}[2][1.0]{
  \dimendef\grafheight=255\dimendef\grafvshift=254
  \setbox0 = \hbox{\scalebox{#1}{\includegraphics{images/#2}}}
  \grafheight=\the\ht0
  \grafvshift=-0.5\grafheight
  \advance\grafvshift by 0.5ex
  \raisebox{\grafvshift}{\includegraphics[height=\grafheight]{images/#2}\xspace}
}

\usepackage{latexsym}
\usepackage{amssymb}
\expandafter\let\csname equation*\endcsname\relax
\expandafter\let\csname endequation*\endcsname\relax
\usepackage{amsmath} 
\usepackage{stmaryrd}
\usepackage{mathtools}
\usepackage{stackrel}

\usepackage{amsthm}
\newtheorem{theorem}{Theorem}[section]

\newtheorem{lemma}[theorem]{Lemma}

\theoremstyle{definition}
\theoremstyle{definition}
\theoremstyle{definition}\newtheorem{definition}[theorem]{Definition}
\theoremstyle{definition}
\theoremstyle{definition}
\theoremstyle{definition}

\DeclareMathOperator{\argmax}{argmax}

\newcommand{\bra}[1]{
    \ensuremath{\left\langle #1 \right|}\xspace}
\newcommand{\ket}[1]{
    \ensuremath{\left|  #1 \right\rangle}\xspace}

\newcommand{\CX}{\ensuremath{\mathrm{CNOT}}\xspace}

\newcommand{\CNOT}{\CX}
\newcommand{\CNOTs}{\CNOT{}s\xspace}

\ifx\numericids\undefined

\else

\fi

\newcommand{\zxcalculus}{\textsc{zx}-calculus\xspace}
\newcommand{\zxdiagram}{\textsc{zx}-diagram\xspace}

\usepackage{dsfont}

\newcommand{\tket}{\ensuremath{\mathsf{t}|\mathsf{ket}\rangle}\xspace}

\usepackage{subcaption}
\usepackage{csvsimple}
\usepackage{multirow}
\usepackage{makecell}
\usepackage{texlogos} 
\usepackage{algpseudocode}
\usepackage{ragged2e}

\DeclareFixedFont{\ttb}{T1}{txtt}{bx}{n}{10} 
\DeclareFixedFont{\ttm}{T1}{txtt}{m}{n}{10}  

\usepackage{color}
\definecolor{deepblue}{rgb}{0,0,0.5}
\definecolor{deepred}{rgb}{0.6,0,0}
\definecolor{deepgreen}{rgb}{0,0.5,0}

\usepackage{listings}
\newcommand\pythonstyle{\lstset{
language=Python,
basicstyle=\ttm,
otherkeywords={self},             
keywordstyle=\ttb\color{deepblue},
emph={MyClass,__init__},          
emphstyle=\ttb\color{deepred},    
stringstyle=\color{deepgreen},
frame=tb,                         
numbers=left,
numberstyle=\tiny\color{gray},
showstringspaces=false            
xleftmargin=10pt,
xrightmargin=10pt,
}}

\lstnewenvironment{python}[1][]
{
\pythonstyle
\lstset{#1}
}
{}

\newcommand\pythoninline[1]{{\pythonstyle\lstinline!#1!}}

\usepackage{hyperref}

%% file: main.bbl
\begin{thebibliography}{10}
\providecommand{\bibitemdeclare}[2]{}
\providecommand{\surnamestart}{}
\providecommand{\surnameend}{}
\providecommand{\urlprefix}{Available at }
\providecommand{\url}[1]{\texttt{#1}}
\providecommand{\href}[2]{\texttt{#2}}
\providecommand{\urlalt}[2]{\href{#1}{#2}}
\providecommand{\doi}[1]{doi:\urlalt{http://dx.doi.org/#1}{#1}}
\providecommand{\bibinfo}[2]{#2}

\bibitemdeclare{article}{Amy_2018}
\bibitem{Amy_2018}
\bibinfo{author}{Matthew \surnamestart Amy\surnameend},
  \bibinfo{author}{Parsiad \surnamestart Azimzadeh\surnameend} \&
  \bibinfo{author}{Michele \surnamestart Mosca\surnameend}
  (\bibinfo{year}{2018}): \emph{\bibinfo{title}{On the controlled-NOT
  complexity of controlled-NOT–phase circuits}}.
\newblock {\sl \bibinfo{journal}{Quantum Science and Technology}}
  \bibinfo{volume}{4}(\bibinfo{number}{1}), p. \bibinfo{pages}{015002},
  \doi{10.1088/2058-9565/aad8ca}.

\bibitemdeclare{article}{staq}
\bibitem{staq}
\bibinfo{author}{Matthew \surnamestart Amy\surnameend} \& \bibinfo{author}{Vlad
  \surnamestart Gheorghiu\surnameend} (\bibinfo{year}{2020}):
  \emph{\bibinfo{title}{staq—A full-stack quantum processing toolkit}}.
\newblock {\sl \bibinfo{journal}{Quantum Science and Technology}}
  \bibinfo{volume}{5}(\bibinfo{number}{3}), p. \bibinfo{pages}{034016},
  \doi{10.1088/2058-9565/ab9359}.

\bibitemdeclare{article}{Amy2014Polynomial-Time}
\bibitem{Amy2014Polynomial-Time}
\bibinfo{author}{Matthew \surnamestart Amy\surnameend}, \bibinfo{author}{Dmitri
  \surnamestart Maslov\surnameend} \& \bibinfo{author}{Michele \surnamestart
  Mosca\surnameend} (\bibinfo{year}{2014}):
  \emph{\bibinfo{title}{Polynomial-Time T-Depth Optimization of Clifford+T
  Circuits Via Matroid Partitioning}}.
\newblock {\sl \bibinfo{journal}{IEEE Transactions on Computer-Aided Design of
  Integrated Circuits and Systems}} \bibinfo{volume}{33}(\bibinfo{number}{10}),
  pp. \bibinfo{pages}{1476--1489}, \doi{10.1109/TCAD.2014.2341953}.

\bibitemdeclare{article}{Coecke:2009aa}
\bibitem{Coecke:2009aa}
\bibinfo{author}{Bob \surnamestart Coecke\surnameend} (\bibinfo{year}{2010}):
  \emph{\bibinfo{title}{Quantum picturalism}}.
\newblock {\sl \bibinfo{journal}{Contemporary Physics}}
  \bibinfo{volume}{51}(\bibinfo{number}{1}), pp. \bibinfo{pages}{59--83},
  \doi{10.1080/00107510903257624}.

\bibitemdeclare{book}{Coecke2017Picturing-Quant}
\bibitem{Coecke2017Picturing-Quant}
\bibinfo{author}{Bob \surnamestart Coecke\surnameend} \& \bibinfo{author}{Aleks
  \surnamestart Kissinger\surnameend} (\bibinfo{year}{2017}):
  \emph{\bibinfo{title}{Picturing Quantum Processes: A First Course in Quantum
  Theory and Diagrammatic Reasoning}}.
\newblock \bibinfo{publisher}{Cambridge University Press},
  \doi{10.1017/9781316219317}.

\bibitemdeclare{inproceedings}{Alexander-Cowtan:2019aa}
\bibitem{Alexander-Cowtan:2019aa}
\bibinfo{author}{Alexander \surnamestart Cowtan\surnameend},
  \bibinfo{author}{Silas \surnamestart Dilkes\surnameend},
  \bibinfo{author}{Ross \surnamestart Duncan\surnameend},
  \bibinfo{author}{Alexandre \surnamestart Krajenbrink\surnameend},
  \bibinfo{author}{Will \surnamestart Simmons\surnameend} \&
  \bibinfo{author}{Seyon \surnamestart Sivarajah\surnameend}
  (\bibinfo{year}{2019}): \emph{\bibinfo{title}{{On the Qubit Routing
  Problem}}}.
\newblock In \bibinfo{editor}{Wim \surnamestart van Dam\surnameend} \&
  \bibinfo{editor}{Laura \surnamestart Mancinska\surnameend}, editors: {\sl
  \bibinfo{booktitle}{14th Conference on the Theory of Quantum Computation,
  Communication and Cryptography (TQC 2019)}}, {\sl \bibinfo{series}{Leibniz
  International Proceedings in Informatics (LIPIcs)}} \bibinfo{volume}{135},
  \bibinfo{publisher}{Schloss Dagstuhl--Leibniz-Zentrum fuer Informatik},
  \bibinfo{address}{Dagstuhl, Germany}, pp. \bibinfo{pages}{5:1--5:32},
  \doi{10.4230/LIPIcs.TQC.2019.5}.
\newblock \urlprefix\url{http://drops.dagstuhl.de/opus/volltexte/2019/10397}.

\bibitemdeclare{article}{Cowtan:2019aa}
\bibitem{Cowtan:2019aa}
\bibinfo{author}{Alexander \surnamestart Cowtan\surnameend},
  \bibinfo{author}{Silas \surnamestart Dilkes\surnameend},
  \bibinfo{author}{Ross \surnamestart Duncan\surnameend}, \bibinfo{author}{Will
  \surnamestart Simmons\surnameend} \& \bibinfo{author}{Seyon \surnamestart
  Sivarajah\surnameend} (\bibinfo{year}{2020}): \emph{\bibinfo{title}{Phase
  Gadget Synthesis for Shallow Circuits}}.
\newblock {\sl \bibinfo{journal}{Electronic Proceedings in Theoretical Computer
  Science}} \bibinfo{volume}{318}, pp. \bibinfo{pages}{213--228},
  \doi{10.4204/eptcs.318.13}.

\bibitemdeclare{article}{Dawson:2005aa}
\bibitem{Dawson:2005aa}
\bibinfo{author}{Christopher~M. \surnamestart Dawson\surnameend},
  \bibinfo{author}{Andrew~P. \surnamestart Hines\surnameend},
  \bibinfo{author}{Duncan \surnamestart Mortimer\surnameend},
  \bibinfo{author}{Henry~L. \surnamestart Haselgrove\surnameend},
  \bibinfo{author}{Michael~A. \surnamestart Nielsen\surnameend} \&
  \bibinfo{author}{Tobias~J. \surnamestart Osborne\surnameend}
  (\bibinfo{year}{2005}): \emph{\bibinfo{title}{Quantum Computing and
  Polynomial Equations over the Finite Field Z2}}.
\newblock {\sl \bibinfo{journal}{Quantum Info. Comput.}}
  \bibinfo{volume}{5}(\bibinfo{number}{2}), p. \bibinfo{pages}{102–112},
  \doi{10.26421/QIC5.2-2}.

\bibitemdeclare{misc}{gray}
\bibitem{gray}
\bibinfo{author}{Frank \surnamestart Gray\surnameend} (\bibinfo{year}{1953}):
  \emph{\bibinfo{title}{Pulse Code Communication}}.

\bibitemdeclare{inbook}{steinercompl}
\bibitem{steinercompl}
\bibinfo{author}{Richard~M. \surnamestart Karp\surnameend}
  (\bibinfo{year}{2010}): \emph{\bibinfo{title}{Reducibility Among
  Combinatorial Problems}}, pp. \bibinfo{pages}{219--241}.
\newblock \bibinfo{publisher}{Springer Berlin Heidelberg},
  \bibinfo{address}{Berlin, Heidelberg}, \doi{10.1007/978-3-540-68279-0\_8}.

\bibitemdeclare{article}{Kissinger:2019ac}
\bibitem{Kissinger:2019ac}
\bibinfo{author}{Aleks \surnamestart Kissinger\surnameend} \&
  \bibinfo{author}{Arianne \surnamestart {Meijer-van de Griend}\surnameend}
  (\bibinfo{year}{2020}): \emph{\bibinfo{title}{Cnot circuit extraction for
  topologically-constrained quantum memories}}.
\newblock {\sl \bibinfo{journal}{Quantum information \& computation}}
  \bibinfo{volume}{20}(\bibinfo{number}{7-8}), pp. \bibinfo{pages}{581--596},
  \doi{10.26421/QIC20.7-8-4}.

\bibitemdeclare{article}{patel}
\bibitem{patel}
\bibinfo{author}{Patel \surnamestart K.N.\surnameend}, \bibinfo{author}{Markov
  \surnamestart I.L.\surnameend} \& \bibinfo{author}{Hayes \surnamestart
  J.P.\surnameend} (\bibinfo{year}{2008}): \emph{\bibinfo{title}{Optimal
  synthesis of linear reversible circuits}}.
\newblock {\sl \bibinfo{journal}{Quantum Information and Computation}}
  \bibinfo{volume}{8}(\bibinfo{number}{3\&4}), pp. \bibinfo{pages}{282--294},
  \doi{10.26421/qic8.3-4-4}.
\newblock \urlprefix\url{https://cir.nii.ac.jp/crid/1360294647045719424}.

\bibitemdeclare{article}{Nash:2019aa}
\bibitem{Nash:2019aa}
\bibinfo{author}{Beatrice \surnamestart Nash\surnameend}, \bibinfo{author}{Vlad
  \surnamestart Gheorghiu\surnameend} \& \bibinfo{author}{Michele \surnamestart
  Mosca\surnameend} (\bibinfo{year}{2020}): \emph{\bibinfo{title}{Quantum
  circuit optimizations for NISQ architectures}}.
\newblock {\sl \bibinfo{journal}{Quantum Science and Technology}}
  \bibinfo{volume}{5}(\bibinfo{number}{2}), p. \bibinfo{pages}{025010},
  \doi{10.1088/2058-9565/ab79b1}.

\bibitemdeclare{article}{Preskill2018quantumcomputingin}
\bibitem{Preskill2018quantumcomputingin}
\bibinfo{author}{John \surnamestart Preskill\surnameend}
  (\bibinfo{year}{2018}): \emph{\bibinfo{title}{Quantum {C}omputing in the
  {NISQ} era and beyond}}.
\newblock {\sl \bibinfo{journal}{{Quantum}}} \bibinfo{volume}{2},
  p.~\bibinfo{pages}{79}, \doi{10.22331/q-2018-08-06-79}.

\bibitemdeclare{article}{TKETPAPERHERE}
\bibitem{TKETPAPERHERE}
\bibinfo{author}{Seyon \surnamestart Sivarajah\surnameend},
  \bibinfo{author}{Silas \surnamestart Dilkes\surnameend},
  \bibinfo{author}{Alexander \surnamestart Cowtan\surnameend},
  \bibinfo{author}{Will \surnamestart Simmons\surnameend},
  \bibinfo{author}{Alec \surnamestart Edgington\surnameend} \&
  \bibinfo{author}{Ross \surnamestart Duncan\surnameend}
  (\bibinfo{year}{2021}): \emph{\bibinfo{title}{T$|$ket$\rangle $: A
  Retargetable Compiler for {{NISQ}} Devices}}.
\newblock {\sl \bibinfo{journal}{Quantum Science and Technology}}
  \bibinfo{volume}{6}(\bibinfo{number}{1}), p. \bibinfo{pages}{014003},
  \doi{10.1088/2058-9565/ab8e92}.

\bibitemdeclare{article}{Soeken:2019aa}
\bibitem{Soeken:2019aa}
\bibinfo{author}{Mathias \surnamestart Soeken\surnameend},
  \bibinfo{author}{Giulia \surnamestart Meuli\surnameend},
  \bibinfo{author}{Bruno \surnamestart Schmitt\surnameend},
  \bibinfo{author}{Fereshte \surnamestart Mozafari\surnameend},
  \bibinfo{author}{Heinz \surnamestart Riener\surnameend} \&
  \bibinfo{author}{Giovanni \surnamestart De~Micheli\surnameend}
  (\bibinfo{year}{2020}): \emph{\bibinfo{title}{Boolean satisfiability in
  quantum compilation}}.
\newblock {\sl \bibinfo{journal}{Philosophical Transactions Of The Royal
  Society A-Mathematical Physical And Engineering Sciences}}
  \bibinfo{volume}{378}(\bibinfo{number}{2164}), p. \bibinfo{pages}{20190161},
  \doi{10.1098/rsta.2019.0161}.
\newblock \urlprefix\url{http://infoscience.epfl.ch/record/275628}.

\bibitemdeclare{inproceedings}{Wille:2019aa}
\bibitem{Wille:2019aa}
\bibinfo{author}{Robert \surnamestart Wille\surnameend}, \bibinfo{author}{Lukas
  \surnamestart Burgholzer\surnameend} \& \bibinfo{author}{Alwin \surnamestart
  Zulehner\surnameend} (\bibinfo{year}{2019}): \emph{\bibinfo{title}{Mapping
  Quantum Circuits to IBM QX Architectures Using the Minimal Number of SWAP and
  H Operations}}.
\newblock In: {\sl \bibinfo{booktitle}{Proceedings of the 56th Annual Design
  Automation Conference 2019}}, \bibinfo{series}{DAC '19},
  \bibinfo{publisher}{Association for Computing Machinery},
  \bibinfo{address}{New York, NY, USA}, pp. \bibinfo{pages}{1--6},
  \doi{10.1145/3316781.3317859}.

\bibitemdeclare{article}{wu2019optimization}
\bibitem{wu2019optimization}
\bibinfo{author}{Bujiao \surnamestart Wu\surnameend}, \bibinfo{author}{Xiaoyu
  \surnamestart He\surnameend}, \bibinfo{author}{Shuai \surnamestart
  Yang\surnameend}, \bibinfo{author}{Lifu \surnamestart Shou\surnameend},
  \bibinfo{author}{Guojing \surnamestart Tian\surnameend},
  \bibinfo{author}{Jialin \surnamestart Zhang\surnameend} \&
  \bibinfo{author}{Xiaoming \surnamestart Sun\surnameend}
  (\bibinfo{year}{2023}): \emph{\bibinfo{title}{Optimization of CNOT circuits
  on limited-connectivity architecture}}.
\newblock {\sl \bibinfo{journal}{Phys. Rev. Res.}} \bibinfo{volume}{5}, p.
  \bibinfo{pages}{013065}, \doi{10.1103/PhysRevResearch.5.013065}.

\bibitemdeclare{article}{Zulehner:2017aa}
\bibitem{Zulehner:2017aa}
\bibinfo{author}{Alwin \surnamestart Zulehner\surnameend},
  \bibinfo{author}{Alexandru \surnamestart Paler\surnameend} \&
  \bibinfo{author}{Robert \surnamestart Wille\surnameend}
  (\bibinfo{year}{2019}): \emph{\bibinfo{title}{An Efficient Methodology for
  Mapping Quantum Circuits to the IBM QX Architectures}}.
\newblock {\sl \bibinfo{journal}{IEEE Transactions on Computer-Aided Design of
  Integrated Circuits and Systems}} \bibinfo{volume}{38}(\bibinfo{number}{7}),
  pp. \bibinfo{pages}{1226--1236}, \doi{10.1109/TCAD.2018.2846658}.

\end{thebibliography}
